\documentclass[12pt]{article}
\usepackage{amssymb,amsmath,amscd,epsfig}
\usepackage{psfrag}
\makeatletter

\@addtoreset{equation}{section}
\def\section{\@startsection {section}{1}{\z@}{-2.25ex plus -1ex minus
 -.2ex}{1.0ex plus .2ex}{\large\bf}}
\def\subsection{\@startsection{subsection}{2}{\z@}{-2.0ex plus%
 -1ex minus -.2ex}{0.5ex plus .2ex}{\bf}}

\def\Ad{\mathrm{Ad}}
\def\ad{\mathrm{ad}}

\newcommand{\inv}[0]{{-1}}

\newcommand{\cif}[0]{\mathcal{C}^\infty}

\newcommand{\oo}[0]{\otimes}

\newcommand{\la}{\langle}
\newcommand{\ra}{\rangle}

\def\bj{{\mbox{\boldmath $j$}}}

\def\bp{{\mbox{\boldmath $p$}}}

\def\bq{{\mbox{\boldmath $q$}}}

\newcommand{\NN}{\mathbb{N}}
\newcommand{\ZZ}{\mathbb{Z}}
\newcommand{\QQ}{\mathbb{Q}}
\newcommand{\RR}{\mathbb{R}}
\newcommand{\CC}{\mathbb{C}}

\newtheorem{theorem}{Theorem}[section]
\newtheorem{lemma}[theorem]{Lemma}
\newtheorem{corollary}[theorem]{Corollary}

\newtheorem{definition}[theorem]{Definition}

\def\bmz{\left(\begin{array}{2,2}}
\def\emz{\end{array}\right)}
\def\bmd{\left(\begin{array}{3,3}}
\def\emd{\end{array}\right)}
\newcommand{\be}{\begin{equation}}
\newcommand{\ee}{\end{equation}}
\newcommand{\ba}{\begin{eqnarray}}
\newcommand{\ea}{\end{eqnarray}}

\def\bpm{\begin{pmatrix}}
\def\epm{\end{pmatrix}}

\begin{document}
\parskip 6pt
\parindent 0pt
\begin{flushright}
\end{flushright}

\begin{center}
\baselineskip 24 pt {\large \bf  Combinatorial quantisation of \\
the Euclidean torus universe}

\baselineskip 16 pt

\vspace{.7cm} {{ C.~Meusburger}\footnote{\tt  catherine.meusburger@uni-hamburg.de}\\
Fachbereich Mathematik,
Universit\"at Hamburg\\
Bundesstra\ss e 55,
D-20146 Hamburg, Germany\\

\vspace{.5cm}
{ K.~Noui}\footnote{\tt karim.noui@lmpt.univ-tours.fr} \\
Laboratoire de Math\'ematiques et de Physique Th\'eorique \\
F\'ed\'eration Denis Poisson Orl\'eans-Tours, CNRS/UMR 6083 \\
Facult\'e des Sciences, Parc de Grammont, 37200 Tours, France} \\

\vspace{0.5cm}

{27 July 2010}

\end{center}

\begin{abstract}
\noindent  
We quantise the Euclidean torus universe  via a combinatorial quantisation formalism based on its formulation as a Chern-Simons gauge theory and on the representation theory of the Drinfel'd double $DSU(2)$. The resulting quantum algebra of  observables is given by  two commuting copies of the Heisenberg algebra, and the associated  Hilbert space can be identified with the space of square integrable functions on the torus. We show that this Hilbert space carries a unitary representation of the modular group and discuss the role of modular invariance in the theory. 
We derive the classical limit of the theory and 
 relate the quantum observables to the geometry of the torus universe.
 \end{abstract}


\section{Introduction}
\label{intro}

Three-dimensional (3d) gravity plays an important role as a toy model for the quantisation of gravity. As the theory simplifies considerably in three dimensions, it allows one to investigate conceptual questions of quantum gravity in a  fully and rigourously quantised theory  \cite{Carlipbook}. Due to its rich mathematical structure, three-dimensional (quantum) gravity  has also attracted strong interest in mathematics.
Major progress was triggered by  the discovery \cite{AT, Witten1} that the theory can be formulated as a Chern-Simons gauge theory  which allowed one to employ gauge theoretical concepts and methods  in its quantisation. In particular, it  related  three-dimensional gravity to the
theory of link and knot invariants \cite{Witten2, restu} and
established the  role of quantum groups and ribbon categories in its quantisation.  These structures  are well-understood in the quantisation of Chern-Simons theories with compact gauge groups.

However, in the context of quantum gravity, it is the non-compact case that is  of special relevance. The gauge groups arising in the Chern-Simons formulation of 3d gravity are the isometry groups of  Euclidean or Lorentzian spacetimes of  constant curvature. For Lorentzian signature, these spacetimes are 3d Minkowski, de Sitter and anti-de Sitter space with isometry groups $ISO(2,1)$, $SO(3,1)$ and $SO(2,1)\times SO(2,1)$. In the Euclidean case, they are three-dimensional Euclidean space, the three-sphere and three-dimensional hyperbolic space with  isometry groups $ISO(3)$, $SO(3)\times SO(3)$ and $SO(3,1)$. 
With the exception of $SO(3)\times SO(3)$, all of these Lie groups are non-compact. 
The representation theory of the  associated quantum groups presents considerable complications compared to the compact case,  and little is known about the associated character and representation rings. 

These difficulties associated with the non-compact case are especially apparent in (discrete) path integral formalisms or state sum approaches to quantisation, where they give rise to ill-defined expressions and divergences that require regularisation. 
Although these issues have been addressed in specific cases such as the Ponzano-Regge model \cite{barnai}, their resolution is still subject of current research 
 - see for instance the recent work \cite{Witten2010} by Witten.  Currently, there is  no coherent framework that allows one to formulate a  consistent path integral quantisation
 for  Chern-Simons theories with non-compact gauge groups such as the ones arising in the Chern-Simons formulation of  three-dimensional gravity.

In this paper, we show how the representation theoretical complications  arising in the quantisation of Chern-Simons theories with non-compact gauge groups  can be resolved
for a simple example,  namely the Euclidean torus universe with vanishing cosmological constant. This corresponds to a Chern-Simons theory with gauge group $ISO(3)$ or $ISU(2)$ 
on a manifold of topology $I\times T$, where $I\subset \RR$ is an interval and $T$ is the torus. 
The relevant quantum group is the Drinfel'd double $DSU(2)$ of the rotation group.  
Although the Drinfel'd double $DSU(2)$ and its representation theory have been investigated in detail  \cite{kM,KBM}, the formulation of a quantum theory based on its representation theory 
faces considerable obstacles due to the fact that its irreducible representations are labelled by a continuous parameter and its representation spaces are infinite-dimensional. 

 For this reason, we do not work with a path integral formalism or a state sum model, but employ a Hamiltonian (canonical) quantisation approach.  
 More specifically, we generalise  the combinatorial quantisation formalism \cite{AGSI,AGSII,AS, BR} for Chern-Simons theories with compact gauge groups, which is rooted in Dirac's quantisation scheme for constrained systems.   
Although it involves the same ingredients  as the path integral approaches - quantum groups and their representation theory -  the central ingredient  of this formalism is not representation theory alone but also an explicit correspondence between Poisson-Lie structures in the classical and quantum groups in the quantum theory.  As we will show in the following, these structures and the close correspondence between the classical and quantum theory allow one to generalise  definitions and results from the  compact to the non-compact setting.

The motivation for considering  manifolds of topology $M\approx I\times T$ is two-fold. Firstly, it allows us to focus on the issues involving the representation theory of the Drinfel'd double $DSU(2)$ without addressing the technical complications that arise from the combinatorics of fundamental groups of higher genus and punctured surfaces. On the other hand, the torus universe is a particularly well-studied example \cite{MD,ARL, giu3,peld1,peld2,carnel,carnel2,carliptor1,carliptor2,carliptor3}, for an overview see \cite{Carlipbook}, which allows one to compare the resulting quantum theory to other quantisation approaches  which do not make use of quantum groups. We expect that our approach can be generalised to manifolds of topology $M\approx I\times \Sigma$, where $\Sigma$ is a surface of general genus $g$ and with $n$ punctures.

The paper is structured as follows. In Sect.~\ref{torusuniverse}, we discuss the  geometry of the flat, Euclidean torus universe and its construction as a quotient of three-dimensional Euclidean space. We summarise its description in terms of a Chern-Simons 
theory with gauge group $ISU(2)$ and the associated parametrisation of phase space and Poisson structure.

In Sect.~\ref{exthilb}, we  explain
 the combinatorial quantisation approach and its generalisation to the Euclidean torus universe. 
We introduce the relevant quantum groups that arise in the quantisation of Euclidean 3d gravity without cosmological constant - the Drinfel'd double $DSU(2)$ and its dual. We then construct the algebra of kinematical (non-gauge invariant) observables, the graph algebra,  and determine its unique irreducible Hilbert space representation, which defines the kinematical Hilbert space of the theory.

Sect.~\ref{physhilb} describes the implementation of the constraints in the quantum theory following the combinatorial quantisation scheme. We define the constraint operators and show that their action on the representation spaces of the graph algebra can be identified with the adjoint representation of $DSU(2)$ on its dual. This allows us to construct the gauge invariant Hilbert space and the algebra of  gauge invariant observables via a suitable regularisation of  the characters of $DSU(2)$. The former  is given by the square integrable functions on the torus, the latter is generated by the  Wilson loop observables associated with the $a$- and $b$-cycles of the torus. 

In Sect.~\ref{btrafo}, we discuss the classical limit of the theory.  We derive an alternative description of the kinematical and gauge invariant Hilbert spaces,  which allows us to directly relate the associated operators to phase space functions in the classical theory. This provides a geometrical interpretation of the quantum theory and demonstrates that the  algebra of gauge invariant quantum observables is given by two commuting copies of the Heisenberg algebra. 

In Sect.~\ref{mapsec}, we investigate the action of  mapping class groups on the kinematical and gauge invariant Hilbert space of the theory and on the associated algebras of observables. We derive an explicit expression for the action of the modular group on the gauge invariant Hilbert space and show that it coincides with the action of certain gauge invariant observables associated with the $a$- and $b$-cycle of the torus. We discuss the role of mapping class group invariance in the quantum theory.   Sect.~\ref{outlook} contains our outlook and conclusions. Appendices \ref{hopfapp} and \ref{su2char} summarise some facts and definitions from, respectively,  the theory of Hopf algebras and from the representation theory of the group $SU(2)$.

\section{The classical Euclidean torus universe}
\label{torusuniverse}

\subsection{Geometry of the torus universe}
\label{classgeom}
In the following, we consider three-dimensional (3d) Euclidean gravity without cosmological constant on manifolds of topology $M\approx I\times T$, where $I\subset \RR$ is an interval and $T$ the torus.  As the Ricci curvature of a three-dimensional manifold determines its sectional curvature, solutions of the vacuum Einstein equations are of constant curvature, which is given by the cosmological constant. This implies that the spacetimes under consideration are 
 flat and locally isometric to three-dimensional Euclidean space $\mathbb E^3$. 
 More specifically, they are obtained as quotients of  $\mathbb E^3$ by the action of a discrete subgroup of $\text{Isom}(\mathbb E^3)=ISO(3)$ which  is  isomorphic to the fundamental group $\pi_1(T)\cong \ZZ\times \ZZ$ of the torus.

The Euclidean group $ISO(3)$ is the semidirect product of the three-dimensional rotation group $SO(3)$ with the abelian group of translations: $ISO(3)=SO(3)\ltimes\RR^3$. Its universal cover is the group $ISU(2)=SU(2)\ltimes \RR^3$. In the following, we parametrise their elements  as 
\begin{align}
\label{param}
(u,\mathbf a)=(u,-\Ad(u)\bj) 
\end{align} where $u\in SO(3)$ or $u\in SU(2)$, $\mathbf a,\bj\in\RR^3$ and  $\Ad$ denotes the adjoint action of $SU(2)$ and $SO(3)$ on $\mathfrak{su}(2)\cong \mathfrak{so}(3)\cong\RR^3$.
 In this parametrisation, the group multiplication laws of $ISO(3)$ and $ISU(2)$  take the form
\begin{align}\label{gmult}
(u_1,\mathbf a_1)\cdot (u_2,\mathbf a_2)=(u_1u_2, \mathbf a_1+\Ad(u_1)\mathbf a_2).
\end{align}
The associated Lie algebra $\mathfrak{iso}(3)\cong\mathfrak{isu}(2)$ is the six-dimensional real Lie algebra spanned by generators $J_a,P_a$, $a=0,1,2$, and with Lie bracket
\begin{align}
\label{liealg} 
[J_a,J_b]=\epsilon_{abc} J^c,\qquad
[J_a,P_b]=\epsilon_{abc} P^c\qquad [P_a,P_b] =0,
\end{align}
where $\epsilon_{abc}$ denotes the totally antisymmetric tensor in three indices with $\epsilon_{012}=1$.
The Lie algebra elements $J_a$ correspond to rotations,  the elements $P_a$ to translations. 
In the  fundamental representation of $\mathfrak{su}(2)$,  the generators $J_a$ are given by the Pauli matrices
$J_a=\tfrac i 2 \sigma_a$, and the exponential map $\exp: \mathfrak{su}(2)\rightarrow SU(2)$ takes the form
\begin{align}
\exp(p^cJ_c)=\cos\tfrac\mu 2 \,1+2 \sin\tfrac \mu 2\, \hat p^cJ_c\quad\text{where}\; \bp\in\RR^3, \bp^2=\mu^2, \hat\bp=\tfrac 1 \mu \bp.
\end{align}
 
The construction of a flat Euclidean spacetime of topology $M\approx I\times T$ as a quotient of Euclidean space $\mathbb E^3$ is given by an injective  group homomorphism $h: \pi_1(T)\rightarrow ISO(3)$ that gives rise to a free and properly discontinuous action of $\pi_1(T)\cong \ZZ\times\ZZ$ on an open region $D\subset \mathbb E^3$. This ensures that the quotient of $\mathbb E^3$ by this group action is a manifold and inherits a flat Euclidean metric from $\mathbb E^3$. 
Group homomorphisms $h,h':\pi_1(T)\rightarrow ISO(3)$ which are related by global conjugation $h'=g\cdot h\cdot g^\inv$ with $g\in ISO(3)$ define isometric and hence physically equivalent torus spacetimes.

\subsubsection{The static torus universe}
\begin{figure}
\includegraphics[scale=0.25]{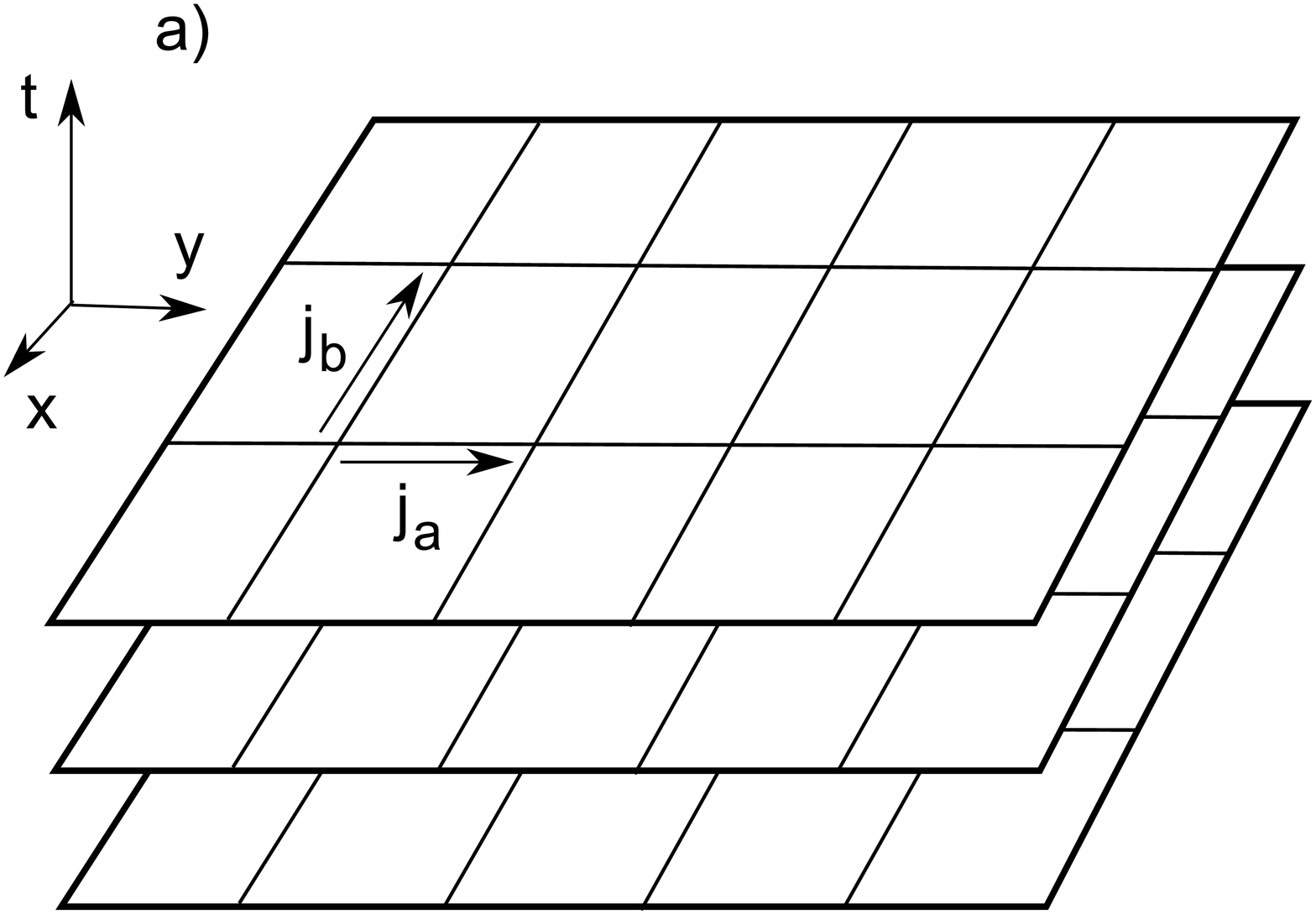}
\includegraphics[scale=0.25]{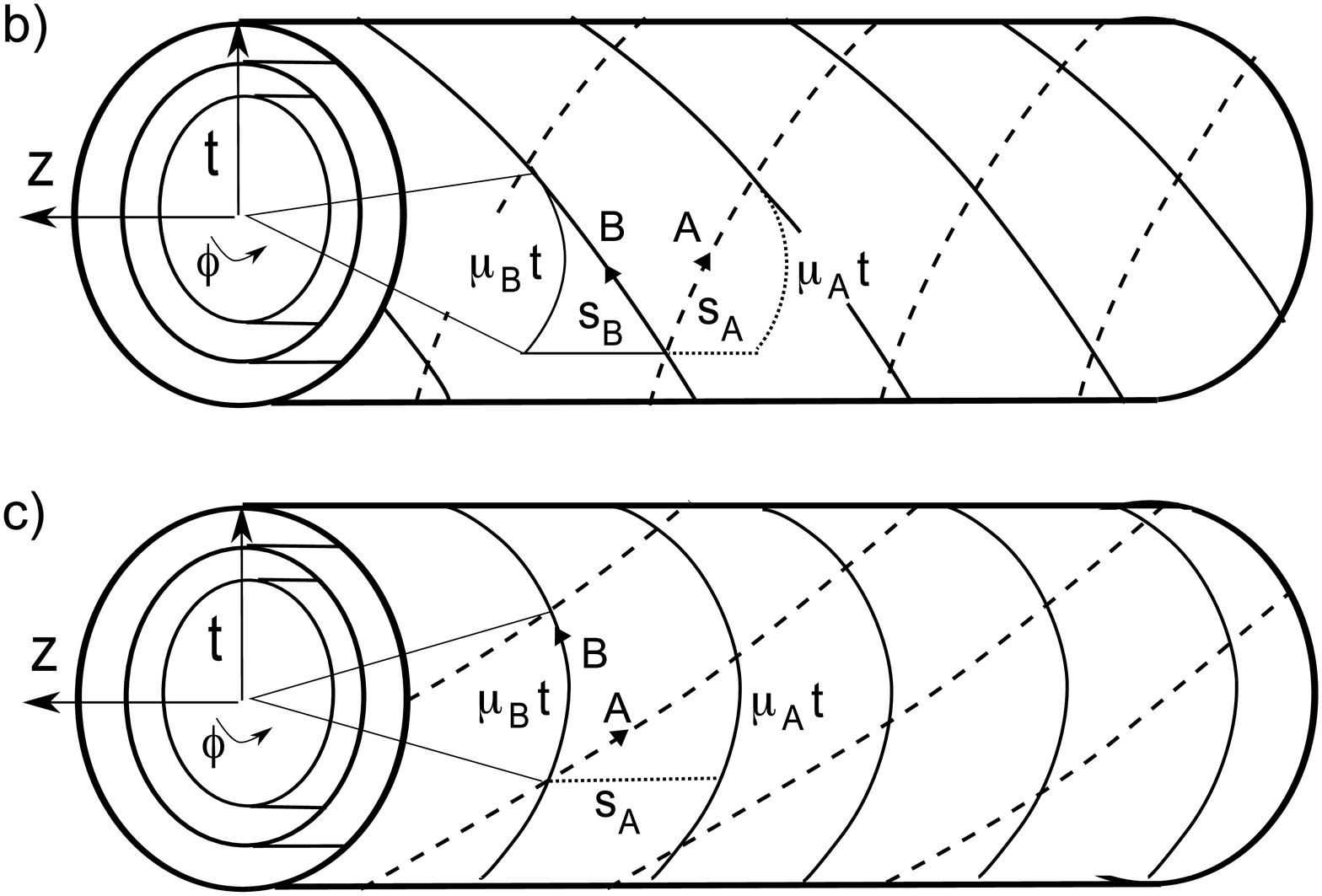}
\caption{The static and the evolving torus universe:\newline  a) Action of $\pi_1(T)$ on $\mathbb E^3$ for the static torus universe ($\mu_A=\mu_B=0$). \newline b) Action of $\pi_1(T)$ on $\mathbb E^3$ for the generic evolving torus universe with $s_A,s_B\neq 0$. \newline c) Action of $\pi_1(T)$ on $\mathbb E^3$ for the evolving torus universe for $s_B=0$.}
\label{torus}
\end{figure}

Injective group homomorphisms $h: \pi_1(T)\rightarrow ISO(3)$  can take two different forms.  The first possibility is that the rotational component of the image vanishes for both the $a$- and the $b$-cycle of the torus. In this case, the images of the $a$ and $b$-cycle 
 are  given by two linearly independent vectors $\bj_A,\bj_B\in\RR^3$
\begin{align}
A=h(a)=(1,\bj_A)\qquad B=h(b)=(1,\bj_B).
\end{align}
The associated action of $\pi_1(T)$ on $\mathbb E^3$ then preserves the affine planes spanned by $\bj_A,\bj_B$. The corresponding torus spacetime is obtained by identifying the points on these planes related by this group action as shown in Figure \ref{torus}.

This implies that the lengths of the $a$- and $b$-cycle and the angle between them are  given by
\begin{align}
l_A=|\bj_A|\qquad l_B=|\bj_B|\qquad \cos\phi_{AB}=\frac{\bj_A\cdot \bj_B}{l_A \cdot l_B} \;.
\end{align}
The fundamental region of the action of $\pi_1(T)$ on the affine plane through $p\in\mathbb E^3$ is the parallelogram $F_p=\{ p  + \lambda_1\bj_A+\lambda_2\bj_B\,|\,\lambda_1,\lambda_2\in[0,1]\}$.
As the length of the $a$-and $b$-cycle and the angle between them do not depend on the coordinate $t$, the associated torus spacetimes are called static.

Because the metric of $\mathbb E^3$ is invariant under rotations, group homomorphisms $h: \ZZ\times\ZZ\rightarrow ISO(3)$ whose rotations components vanish and whose translation components  $\bj_A,\bj_B$ are related by rotations describe the same physical state. 
It is shown in \cite{Carlipbook} that the physically inequivalent static torus spacetimes 
are parametrised  by a complex number, the modulus $\tau\in\CC$, which is given in terms of the variables $\bj_A,\bj_B$ by 
\begin{align}\label{taudef}
\tau=|\tau|e^{i\phi_{AB}}\qquad|\tau|=\frac{l_B}{l_A}\qquad \phi_{AB}=\arccos \left(\frac{\bj_A\cdot \bj_B}{l_A \cdot l_B}\right).
\end{align}
The  physical parameters which characterise  the static torus universe are thus the relative length $|\bj_B|/|\bj_B|$ of the translation vectors $\bj_A,\bj_B$ and the angle between them.

\subsubsection{The evolving torus universe}

In the generic situation, at least one of two rotation components of $h(a)$ and $h(b)$ is non-trivial. In order to commute, the rotation components of $h(a)$ and $h(b)$ must then either  have the same axis or one of them must be trivial. By conjugating them with a suitable translation, one can then impose that
the images $h(a)$, $h(b)$ are of the form
\begin{align}\label{holevol}
A=h(a)=(\exp(-\mu_A\, \hat p^cJ_c), - s_A \hat\bp)\qquad B=h(b)=(\exp(-\mu_B\, \hat p^cJ_c), - s_B\hat\bp),
\end{align}
where $\mu_A,\mu_B\in[0,2\pi]$, $s_A,s_B\in\mathbb R$ and $\hat\bp\in\mathbb R^3$ is a unit vector. Such a group homomorphism gives rise to an action of $\pi_1(T)$ on the 
open region $D=\mathbb E^3\setminus \, \RR\cdot \hat \bp$ obtained by removing the line $\RR\cdot \hat\bp$ from $\mathbb E^3$. This  region is foliated by coaxial  cylinders of radius $t$ 
$$C_t=\{z\hat\bp+ t\hat{\mathbf e}\;|\; z\in\RR, \hat {\mathbf e}\in \RR^3, \hat{\mathbf e}\cdot\hat\bp=0, \hat{\mathbf e}^2=1\},$$  as depicted in Figure \ref{torus}, each of which is preserved under the action of $\pi_1(T)$.
To ensure that the group homomorphism $h:\pi_1(T)\rightarrow ISO(3)$ is injective and the resulting action of $\pi_1(T)$ on the cylinders $C_t$ free and properly discontinuous, one has to impose the additional condition $(\mu_A,s_A)\neq \lambda(\mu_B,s_B)$ for all $\lambda\in\RR$.  If $s_A=0$ ($s_B=0$), one also needs the  to impose $\mu_A=2\pi q$ ($\mu_b=2\pi q$) with $q\in\QQ$.

The corresponding torus spacetime is obtained by identifying on each cylinder $C_t$ the points related by the action of $h(a),h(b)$ as shown in Figure \ref{torus}. By cutting the cylinder $C_t$ along a line parallel to its axis, one obtains a plane tessellated by parallelograms, from which one  deduces the lengths of the $a$- and $b$-cycle and the angle between them
 \begin{align}\label{ldef}
&l_A(t)=\sqrt{s_A^2+t^2 \mu_A^2} \quad l_B(t)=\sqrt{s_B^2+t^2 \mu_B^2}\\
&\phi_{AB}(t)=\arccos\left(\frac{s_As_B+t^2\mu_A\mu_B}{\sqrt{s_A^2+t^2\mu_A^2}\sqrt{s_B^2+t^2\mu_B^2}}\right),\nonumber
\end{align}
 The modulus of the associated torus is given in terms of these quantities by formula \eqref{taudef}. In contrast to the static case, it has an additional dependence on the radial coordinate $t$.  For this reason, the associated torus spacetimes are called evolving.  
 
 Note that these expressions do not depend on the vector $\hat\bp$ in \eqref{holevol}. This is due to the fact that  group homomorphisms $h: \pi_1(T)\rightarrow ISO(3)$ which are related by conjugation with rotations $u\in SO(3)$  define isometric and hence physically equivalent torus spacetimes. The four parameters which characterise physically non-equivalent torus spacetimes are therefore the variables  $\mu_A,\mu_B\in[0,2\pi]$, $s_A, s_B\in\RR$ in \eqref{holevol} or, equivalently, the modulus $\tau(t_0)$ and its derivative $\dot\tau(t_0)$ for a given value $t_0\in\RR^+$.
 Formula \eqref{ldef} provides a direct interpretation of these parameters in terms of the geometry of the torus universe.  In the following, we will refer to the variables $\mu_A,\mu_B$ as masses and to the variables $s_A,s_B$ as spins of the $a$- and $b$-cycle.

\subsection{The description in terms of a Chern-Simons gauge theory}
\label{csdescript}

The geometrical description of the three-dimensional torus universe has been applied very successfully   to the study of various classical aspects and its quantisation \cite{Nelson,LM1,LM2,Matschpart,NP,AA,MD,ARL,AK,TT3D, giu3, carliptor1,carliptor2,carliptor3}.
However, the drawback is that  many of the results obtained are based on the specific geometry of the torus spacetime and cannot easily be generalised to spacetimes of higher genus or with punctures.

As  we intend to use the torus universe as an example for generalising the combinatorial 
quantisation formalism to the non-compact setting, we will focus on its description as a Chern-Simons gauge theory in the following. The Chern-Simons formulation of  3d gravity \cite{AT, Witten1} is obtained from Cartan's formulation of the theory by combining 
the triad $e$ and spin connection $\omega$  into a Chern-Simons gauge field. For Euclidean 3d gravity with vanishing cosmological constant, this gauge field is  $\mathfrak{iso}(3)$-valued and  given by
\begin{align}
\label{gfield}A = \omega_a
J^a + e_a P^a,
\end{align}
where $J_a,P_a$ are the generators of the Lie algebra \eqref{liealg}, $e=e^a_\mu P_a dx^\mu$ is the triad and $\omega=\omega^a_\mu J_a dx^\mu$ the spin connection. 
The associated Chern-Simons action reads
\begin{align}
\label{csact}S_{CS}[A] =\frac{1}{2} \int_M \langle A\wedge  dA\rangle
 +\frac{2}{3}\langle A \wedge A \wedge A\rangle,
\end{align}
where $\langle\,,\rangle$  is the $Ad$-invariant symmetric bilinear form on $\mathfrak{iso}(3)$ defined by the Euclidean metric $\eta=\text{diag}(1,1,1)$
\begin{align}
\label{form}\langle J_a, P_b\rangle = \eta_{ab} \qquad  \langle J_a, J_b\rangle
= \langle P_a,P_b\rangle = 0.
\end{align}
A short computation shows that this action is equivalent to the Einstein-Hilbert action.
The associated equations of motion are a flatness condition on the gauge field
\begin{align}
F_A=dA+ A\wedge A=F_{\omega}^aJ_a+T^aP_a=0,
\end{align}
which combines the requirements of vanishing torsion and curvature of Cartan's formulation
\begin{align}
\label{}T^a=D_\omega e_a = de_a+  \epsilon_{abc}
\omega^b e^c =0\qquad F_{\omega}^a=d\omega^a  + \frac{1}{2} \epsilon^a_{\;bc}
\omega^b\wedge \omega^c=0.
\end{align}
Moreover, it is shown in \cite{Witten1} that the action of infinitesimal diffeomorphisms on triad and spin connection is given by the infinitesimal Chern-Simons gauge transformations.
The two formulations are thus equivalent on the level of action functionals and equations of motion. 

However,  this equivalence is subtle with respect to the choice of the gauge group and the non-degeneracy of the metric. Firstly, the correspondence between Chern-Simons theory and 3d gravity involves only the $\mathfrak{iso}(3)$-valued gauge field and therefore  determines the gauge group of the associated Chern-Simons theory only up to the choice of a cover. For simplicity, we will work with the universal cover $ISU(2)$ in the following.
The second  issue is related to the metric, which  is given in terms of the triad by
\begin{align}
g_{\mu\nu}=\eta_{ab} e^a_\mu e^b_\nu.
\end{align}
 In general relativity, it  is required to be non-degenerate, while no such constraint is imposed in the Chern-Simons formulation. This leads to discrepancies in the global phase space structure of the theory \cite{Maton}, for a discussion in 1+1-dimensions see also \cite{SchS}.

Nevertheless, the  Chern-Simons formulation of 3d gravity has been employed extensively to describe the classical theory and has lead to major progress in its quantisation.
One major advantage is that it relates the phase space of the theory to moduli spaces of flat connections on surfaces. This gives rise to an 
 explicit description of the phase space and Poisson structure in terms of Poisson-Lie groups \cite{FR,AMII},  which serves as a starting point for quantisation. In the case at hand, the phase space of the theory is the moduli space of flat $ISU(2)$ connections on the torus, which is given by
\begin{align}
\mathcal P(T,ISU(2))=\text{Hom}(\ZZ\times\ZZ, ISU(2))/ISU(2).
\end{align}
Note that in contrast to the geometrical formulation, there is no restriction to {\em injective} group homomorphisms $h: \pi_1(T)\rightarrow ISU(2)$, whose image is isomorphic to $\ZZ\times \ZZ$. Consequently, the phase space in the Chern-Simons formulation also contains degenerate group homomorphisms, which do not define a flat Euclidean spacetime of topology $\RR\times T$ as described in Sect.~\ref{classgeom}. This reflects the presence of solutions corresponding  
 to degenerate metrics \cite{Maton}.

\subsection{Fock and Rosly's description of the phase space}
\label{frsect}

\begin{figure}
\centering
\includegraphics[scale=0.4]{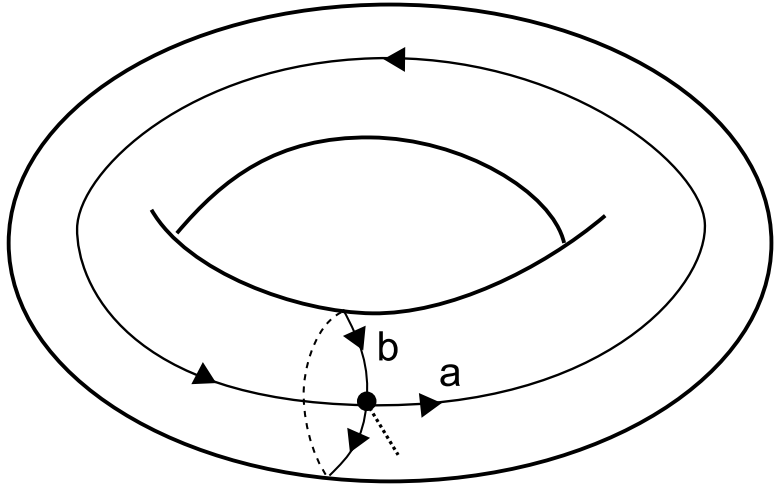}
\caption{Ciliated fat graph associated with a set of generators of the fundamental group of the torus. The dotted line corresponds to the cilium. The lines labelled $a$ and $b$ to the $a$- and $b$-cycle of the torus.}
\label{pi1figure}
\end{figure}

The starting point of the combinatorial quantisation formalism for Chern-Simons gauge theories is Fock and Rosly's description \cite{FR} of their phase space. The phase space of a Chern-Simons theory with gauge group $G$ on a manifold $\RR\times\Sigma$ is the moduli space of flat $G$-connections on $\Sigma$ modulo gauge transformations, which is given as the quotient
\begin{align}
\mathcal P(\Sigma,G)=\text{Hom}(\pi_1(\Sigma), G)/G.
\end{align}
 Fock and Rosly describe the canonical symplectic structure on this space in terms of 
 an auxiliary Poisson structure associated with a ciliated fat graph embedded into $\Sigma$. This is an oriented graph with a linear ordering of the incident edges at each vertex. As the orientation of the surface $\Sigma$ induces a linear ordering of the incident edges at each vertex, a linear ordering of the edges is obtained  by inserting a cilium that separates the incident edges of minimal and maximal order as shown in Figure \ref{pi1figure}.
 
  The simplest ciliated fat graph graph compatible with Fock and Rosly's  description \cite{FR} of the phase space consists of a set of generators of the fundamental group $\pi_1(\Sigma)$. For an oriented surface $\Sigma$ of genus $g\geq 0$ and with $n\geq 0$ punctures, Fock and Rosly's formalism  \cite{FR} defines a Poisson structure on the manifold $G^{n+2g}$.
 The Poisson structure on $\mathcal P(\Sigma,G)$ is then obtained from this auxiliary Poisson structure by imposing a set of $\text{dim}(G)$ first class constraints  which encode the defining relation of the fundamental group $\pi_1(\Sigma)$.

Fock and Rosly's description \cite{FR} not only yields an explicit and simple description of the symplectic structure but also relates it  to well-known Poisson structures from the theory of Poisson Lie groups \cite{FR,AMII}, whose quantisation has been studied extensively, for an overview see \cite{CP}. 
For the case of Euclidean and Lorentzian 3d gravity with vanishing cosmological constant, an explicit description of Fock and Rosly's Poisson structure for a general (punctured) surface $\Sigma$ and a general 
 graph was derived in \cite{MN1}. The description in terms of a set of generators of the fundamental group $\pi_1(\Sigma)$ was first given in 
 \cite{we1,we2,we3}.

  When applied to the torus and the ciliated fat graph depicted 
 in Figure \ref{pi1figure}, 
  this yields a Poisson algebra 
 generated by functions $f\in\cif(SU(2)\times SU(2))$ and by two vectors $\bj_A,\bj_B\in\RR^3$. The former are interpreted as functions of the rotation components of the holonomies along the $a$- and $b$-cycles of the torus. The latter parametrise the translational components of these holonomies as in \eqref{param}:
\begin{align}\label{ABhol}
A\!=\!(u_A,\!-\Ad(u_A)\bj_A)\!=\!(u_A,0)\cdot(e,\!-\bj_A)\quad B\!=\!(u_B,\!-\Ad(u_B)\bj_B)\!=\!(u_B,0)\cdot(e,\!-\bj_B).
\end{align}
In terms of these variables, Fock and Rosly's Poisson structure \cite{FR} takes the form
\begin{align}
\label{poissjj}&\{j^a_A,j^b_A\}=-\epsilon^{ab}_{\;\;\;\;c}\, j^c_A\qquad\{j^a_B,j^b_B\}=-\epsilon^{ab}_{\;\;\;\;c} \,j^c_B\qquad \{j^a_A,j^b_B\}=-\epsilon^{ab}_{\;\;\;\;c}\,j^c_{B}\\
&\{j^a_A, f\}(u_A,u_B)=- \frac d {dt}\bigg |_{t=0} f(e^{-tJ_a} u_A e^{tJ_a}\,,\, [e^{-tJ_a}, u_A]u_B e^{tJ_a})\label{poissfj}\\
&\{j^a_B, f\}(u_A,u_B)=-\frac d {dt}\bigg|_{t=0} f(e^{-tJ_a}u_A, e^{-tJ_a}u_B e^{tJ_a})\nonumber\\
\label{poissj}
&\{f,g\}(u_A,u_B)=0\qquad\qquad \qquad\qquad\qquad \qquad\forall f,g\in\cif(SU(2)\times SU(2)).
\end{align}
This implies that the  components of the vectors $\bj_A,\bj_B$ can  be identified with  certain vector fields on $SU(2)\times SU(2)$. Their Poisson brackets \eqref{poissjj} then coincide with the Lie brackets of these vector fields and their Poisson brackets \eqref{poissfj} with functions  $f\in\cif(SU(2)\times SU(2))$ are given by the action of these vector fields on functions. 

The requirement that the group elements $A,B\in ISU(2)$ are the images of a the $a$- and $b$-cycle under a group homomorphism  $h:\pi_1(T)\cong\ZZ\times\ZZ\rightarrow ISU(2)$  results in a constraint on the variables $A,B$ which encodes the defining relation  of the fundamental group $\pi_1(T)$
\begin{align}\label{cdef}
C=(u_C, -\Ad(u_C)\bj_C)=A^\inv \cdot B\cdot A\cdot B^\inv\approx 1. \end{align}
This $ISU(2)$-valued constraint defines a set of six first class constraints in the terminology of Dirac \cite{HT}. The associated gauge transformations are given by the Poisson brackets
 of the constraint variables $\bj_C, g_C=g(u_C)$ with the components of the vectors $\bj_A,\bj_B$ and with functions $f\in\cif(SU(2)\times SU(2))$
\begin{align}\label{constcl}
&\{j^a_C, j^b_X\}=-\epsilon^{ab}_{\;\;\;d}\,j^d_X\\
&\{j^a_C, f\}(u_A,u_B)=- \frac d {dt}\bigg |_{t=0}f(e^{-tJ_a} u_A e^{tJ_a}, e^{-tJ_a}u_B e^{tJ_a} )\nonumber\\
&\{j^a_X, g_C\}(u_A,u_B)=-\frac{d}{dt}\bigg|_{t=0} g(u_C\cdot[u_X, e^{-tJ_a}]),\nonumber
\end{align}
where $X=A$ or $B$, $g\in\cif(SU(2))$ and $g_C(u_A,u_B):=g([u_A^\inv, u_B])$. 

The physical (gauge invariant) observables of Fock and Rosly's Poisson algebra are elements that Poisson commute with the action of the constraint in \eqref{constcl}. A specific set of such observables is provided by the Wilson loops associated with closed curves on the punctured torus and the two quadratic Casimir operators of $\mathfrak{iso}(3)$. For Euclidean 3d gravity, these Wilson loops are studied in detail in \cite{MN1, we1,we2,we3,ich,ich3}. In terms of the parametrisation 
\begin{align}
X=(\exp(p^c_XJ_c), -\Ad(\exp(p^c_XJ_c))\bj_X) \in ISU(2)\end{align}
they take the form
\begin{align}\label{ob}
\mu_X=\sqrt{\bp^2_X}\qquad s_X=\bj_X\bp_X/\mu_X.
\end{align}
The  two fundamental Wilson loop observables associated with the $a$- and $b$-cycle are thus directly related to the mass and spin observables $\mu_A$, $\mu_B$, $s_A$, $s_B$ defined in the previous subsection, which encode the geometry of the torus universe.

\subsection{Quantisation and constraint implementation}
\label{constattempt}

The quantisation of Fock and Rosly's Poisson algebra \cite{FR} for a Chern-Simons theory with gauge group $G=ISU(2)$ on a genus $g\geq 0$ surface $\Sigma$  with $n\geq 0$ punctures and a for general ciliated fat graph is  constructed in \cite{MN1}. The case of a ciliated fat graph consisting of set of generators of the fundamental group $\pi_1(\Sigma)$ is treated in  \cite{we2, we3}. When restricted to the case at hand, the results of \cite{MN1,we2,we3} state that that the associated quantum algebra is generated by six operators $\hat\j_A^a,\hat \j_B^a$, $a=0,1,2$, which are the quantum counterparts of the  vectors $\bj_A,\bj_B$, and by functions $f\in\cif(SU(2)\times SU(2))$. It is shown in \cite{we2} that this algebra has a single unitary irreducible representation 
on  the Hilbert space $\mathcal H_{kin}=L^2(SU(2)\times SU(2))$ with the standard inner product induced by the Haar measure on $SU(2)$. Its action on this Hilbert space is given by
\begin{align}\label{jfact}
&\Pi(\hat \j^a_A)\psi(u_A,u_B)=\{j^a_A, \psi\}(u_A,u_B)=- \frac d {dt}\bigg |_{t=0} \psi(e^{-tJ_a} u_A e^{tJ_a}\,,\, [e^{-tJ_a}, u_A] u_B e^{tJ_a})\nonumber\\
& \Pi(\hat \j^a_B)\psi(u_A,u_B)=\{j^a_B,\psi\}(u_A,u_B)=-\frac d {dt}\bigg|_{t=0} \psi(e^{-tJ_a} u_A\,,\, e^{-tJ_a} u_B e^{tJ_a})\nonumber\\
&\Pi(f)\psi(u_A,u_B)=f(u_A,u_B)\cdot \psi(u_A,u_B),
\end{align}
for $\psi\in L^2(SU(2)\times SU(2))$. In particular, the quantum counterparts of the mass and spin variables $\mu_A$, $\mu_B$, $s_A$, $s_B$  act on the states $\psi$ according to
\begin{align}\label{masspinact}
&\Pi(\hat \mu_A)\psi(u_A,u_B)=\mu(u_A)\cdot \psi(u_A,u_B) & 
&\Pi(\hat s_A)\psi(u_A,u_B)=\frac d {dt}\bigg|_{t=0}\!\!\!\!\psi(u_A, u_B e^{\hat p^a(u_A)J_a})\\
&\Pi(\hat \mu_B)\psi(u_A,u_B)=\mu(u_B)\cdot \psi(u_A,u_B) &
&\Pi(\hat s_B)\psi(u_A,u_B)=\frac d {dt}\bigg|_{t=0}\!\!\!\!\psi( e^{-\hat p^a(u_B)J_a} u_A, u_B)\nonumber,
\end{align}
where $\mu$, $\hat p^a$, $a=0,1,2$, are functions on $SU(2)$ given by the parametrisation of $SU(2)$-elements in terms of  the exponential map: 
\begin{align}\label{mupdef}
\mu(\exp(q^aJ_a))=\sqrt{\bq^2}\in \, [0,2\pi]\qquad \hat p^a(\exp(q^aJ_a))=q^a/{\sqrt{\bq^2}}.\end{align}

The quantum counterparts of the constraint variables  \eqref{cdef} are operators on \linebreak$L^2(SU(2)\times SU(2))$, whose action on states is given by
\begin{align}
&\Pi(g_C)\psi (u_A,u_B)=g([u_A^\inv,u_B])\cdot \psi(u_A,u_B)\label{lconst}\\
&\Pi(\hat \j^a_C)\psi(u_A,u_B)=\frac d {dt}\bigg|_{t=0}\psi(e^{-tJ_a} u_A e^{tJ_a}, e^{-tJ_a}u_Be^{tJ_a}).\label{tconst}
\end{align}
In Dirac's constraint quantisation formalism, the  Hilbert space  of gauge invariant (physical) states is to be defined as the linear subspace  $\mathcal H_{inv}\subset L^2(SU(2)\times SU(2))$ invariant under the action of the constraint operators. Imposing that states are invariant under the action of the constraint operators
 $\hat \j^a_C$, $a=0,1,2$, is straightforward. It amounts to the requirement that they are
  invariant under the diagonal action of $SU(2)$ on $SU(2)\times SU(2)$.
  A basis of such states can be constructed using the characters of $SU(2)$ 
\begin{align}\label{spinnetor}
\psi_{IJK}(g,h)=\int_{SU(2)}\!\!\!\! \!\!\!\! dz\; \chi_I(gz)\chi_J(hz)\chi_K(z),
\end{align}
where $\chi_I,\chi_J,\chi_K$ are the characters of $SU(2)$ given by \eqref{charexpl}. 
This is the technique employed in the loop quantum gravity approach and the state sum approach to three-dimensional gravity, in which these functions are known as spin network states \cite{AK}.  

However, imposing invariance under the  constraint operator \eqref{lconst} turns out to problematic. Requiring that a state $\psi$ is invariant under the action of \eqref{lconst} for all functions $g\in\cif(SU(2)\times SU(2))$
  forces it  to vanish almost everywhere outside the set
  \begin{align}
  N_{AB}=\{(u_A,u_B)\in SU(2)\times SU(2)\;|\; [u_A^\inv,u_B]=1\}.
  \end{align}
  As $N_{AB}\subset SU(2)\times SU(2)$ is a subset of measure null with respect to the Haar measure on $SU(2)\times SU(2)$, there is no  non-trivial state $\psi\in L^2(SU(2)\times SU(2))$ that satisfies this condition. To implement the constraint \eqref{lconst}, one therefore must enlarge the Hilbert space $H_{inv}$ and include distributions on $SU(2)\times SU(2)$.  
  Formally, distributional states which are invariant under the action of the constraint \eqref{lconst} should take the form
\begin{align}\label{naivestate}
\phi(g,h)=\delta_e([g, h])\phi_0(g,h)
\end{align} 
where $\delta_e$ stands for the  delta-distribution on the space of class functions on $SU(2)$ and $\phi_0$ is invariant under simultaneous conjugation of both arguments with $SU(2)$. 

However,
 this naive definition of  gauge invariant states as distributions causes difficulties in the definition of the inner product on the space of gauge invariant states.  This is due to the fact that the canonical inner product on $L^2(SU(2)\times SU(2))$ does not induce an inner product on the space of such distributions.
 When attempting to naively define an inner product using the  inner product $L^2(SU(2)\times SU(2))$, one obtains an expression  involving the product of two delta-distributions
\begin{align}\label{naiveprod}
\langle \psi,\phi\rangle=\int_{SU(2)\times SU(2)} \!\!\!\! \!\!\!\! \!\!\!\! \!\!\!\! dg dh\; \delta_e([g, h])\delta_e([g,h])\overline{ \psi_0(g,h)}\phi_0(g,h),
\end{align}
which is ill-defined and requires regularisation. Further difficulties of this type arise when one attempts to define the action  of gauge invariant observables on the invariant states. 
The  construction of the physical (gauge invariant) Hilbert space with a well-defined inner product and action of gauge invariant observables therefore requires a regularisation, which is difficult to define without additional structures.  Due to these difficulties, previous attempts to construct the gauge invariant Hilbert space associated with three-dimensional gravity with vanishing cosmological constant via Dirac's constraint quantisation scheme have  encountered obstacles and were only partly successful.

For this reason, we employ a different approach. Instead of building on the quantisation of Fock and Rosly's Poisson algebra in \eqref{jfact} and attempting to naively implement the constraint as outlined above, we start from the general theory underlying the combinatorial quantisation of Chern-Simons theory. We then generalise this formalism to the Euclidean torus universe and show how the relevant techniques can be extended to this case by using the representation theory of  the Drinfel'd double $DSU(2)$. 

It turns out that the quantum group symmetries of the theory provide additional structure which yields an explicit definition of the gauge invariant Hilbert space and the algebra of gauge invariant quantum observables. 
 In Sect.~\ref{relpaper}, we will then demonstrate how the description in \cite{MN1,we2} outlined above emerges from this description through a change of basis. In particular, we show that the gauge invariant Hilbert space obtained in the combinatorial formalism defines a regularisation of the ill-defined inner product \eqref{naiveprod}.


\section{Combinatorial quantisation of the Euclidean torus universe}
\label{exthilb}

\subsection{The combinatorial quantisation formalism}

\label{formsumm}

Combinatorial quantisation  is a Hamiltonian quantisation scheme for Chern-Simons gauge theories on manifolds $M\approx\RR\times \Sigma$, where $\Sigma$ is an orientable two-surface of genus $g\geq 0$ with $n\geq 0$ punctures.
It was first proposed and carried out for Chern-Simons theories with compact, semisimple gauge groups by Alekseev, Grosse and Schomerus \cite{AGSI,AGSII,AS}
and by Buffenoir and Roche \cite{BR}. The case of the non-compact  gauge group $SL(2,\CC)$ is treated in \cite{BNR}.  The combinatorial quantisation of Chern-Simons theories with semidirect product gauge groups  is investigated in \cite{we2,we3}. This includes 
 the three-dimensional Poincar\'e and Euclidean group $ISO(2,1)$, $ISO(3)$ arising in the Chern-Simons formulation of 3d gravity with vanishing cosmological constant.
 However,  the publications \cite{we2,we3} focus on the quantisation of Fock and Rosly's Poisson algebra and the mapping class group symmetries in the resulting quantum theory and do not attempt to  construct the gauge invariant Hilbert space.

Before addressing this problem  for the  Euclidean torus universe, we give a brief summary of the combinatorial quantisation formalism.  The starting point of combinatorial quantisation   is Fock and Rosly's description \cite{FR} of the Poisson structure on the moduli space $\mathcal P(\Sigma,G)$ of flat $G$-connections on $\Sigma$. Combinatorial quantisation  can be viewed as an application of  Dirac's constraint quantisation programme to Fock and Rosly's auxiliary Poisson structure \cite{FR} and  proceeds in two steps. The first is the construction of the so-called graph algebra ${\cal L}(\Sigma,G)$ and its irreducible Hilbert space representations. The associative algebra ${\cal L}(\Sigma,G)$ is the quantum counterpart of 
Fock and Rosly's Poisson algebra associated with  a set of generators of the fundamental group $\pi_1(\Sigma)$. 

It has been shown by Alekseev \cite{Alekseev} that the graph algebra is directly related to two algebras from the theory of quantum groups whose representation theory is well-understood.
For  a genus $g$ surface $\Sigma$ with $n$ punctures, one has an algebra isomorphism  $\mathcal L(\Sigma, G)\cong \mathcal A^n\oo H(\mathcal A)^g$, where $\mathcal A$ is a quantum group associated with $G$  and $H(\mathcal A)$ its Heisenberg double algebra.
This isomorphism allows one to directly construct the irreducible Hilbert space representations of ${\cal L}(\Sigma,G)$ from  representations of  the quantum group $\mathcal A$ and its Heisenberg double $H(\mathcal A)$. The resulting representation space of ${\cal L}(\Sigma, G)$ defines the kinematical (non-gauge invariant) Hilbert space of the theory.

 The second step of the combinatorial quantisation formalism is the imposition of the quantum flatness constraint via Dirac's constraint quantisation formalism. The quantum flatness constraint is  associated with an invertible element of the graph algebra $C\in \mathcal L(\Sigma,G)$ and can be  viewed as the quantum counterpart of the defining relation of the fundamental group $\pi_1(\Sigma)$. It induces a set of algebra automorphisms
 of the  graph algebra $\mathcal L(\Sigma,G)$ and acts on its representation spaces. 
 
 The moduli algebra $\mathcal M(\Sigma, G) \subset {\cal L}(\Sigma,G)$ of physical (gauge invariant) observables is  defined as the subalgebra of $\mathcal L(\Sigma,G)$ which is invariant under the algebra automorphisms induced by $C$. It is the quantum counterpart of the algebra of functions on the  moduli space  $\mathcal P(\Sigma,G)$.  The unitary irreducible representations of this algebra 
define the  physical or gauge invariant Hilbert spaces of the theory.  The gauge invariant states which span  this Hilbert space are  obtained from the irreducible representations of the graph algebra $\mathcal L(\Sigma,G)$ by selecting the linear subspace of states that are invariant under the action of $C$.

For Chern-Simons theories with compact semisimple gauge groups $G$, the relevant
quantum groups are $q$-deformed universal enveloping algebras $\mathcal A=U_q(\mathfrak g)$
at roots of unity $q^n=1$, whose representations are well-behaved. Their irreducible representations involve finite-dimensional representation spaces that are labelled by a finite number of discrete parameters and are equipped with a quantum trace. In this case, the gauge invariant  Hilbert space can be constructed using group averaging techniques and the moduli algebra $\mathcal M(\Sigma,G)$ of physical observables is obtained from the quantum trace \cite{AGSI,AGSII,AS, BR}.

 In the case of  non-compact gauge groups, the representation theory of the associated quantum groups and, consequently,  the construction of the moduli algebra $\mathcal M(\Sigma,G)$ and the physical Hilbert space become more involved. This is due to the fact that the irreducible representations of these  quantum groups can involve infinite-dimensional representation spaces  labelled by continuous parameters. This leads to problems in the definition of a quantum trace and  the characters.  For this reason, the techniques used in \cite{AGSI,AGSII,AS,BR} to construct the gauge invariant Hilbert space and the moduli algebra cannot be applied to the non-compact case.  In \cite{BNR} a different technique was developed for the non-compact quantum Lorentz group associated with $SL(2,\mathbb C)$.  In the following, we show how this problem can be addressed for the Euclidean torus universe where the relevant quantum group is the Drinfel'd double $DSU(2)$ of the rotation group $SU(2)$.

\subsection{Quantum group structures for the Euclidean torus universe}

\label{qugroup}

\subsubsection{The Hopf algebra $DSU(2)$ and its dual}  \label{qudouble}

The relevant quantum groups  for the combinatorial quantisation of the Euclidean torus universe
are the Drinfel'd double  $DSU(2)$ and its dual $DSU(2)^*$. The Drinfel'd double $DSU(2)$ is a quasi-triangular ribbon Hopf $*$-algebra  and can be viewed as a  deformation (in the sense of Drinfel'd) of the classical group algebra $\mathbb C(ISU(2))$.  The deformation parameter is related to the Planck length  $\ell_P$, and the deformation involves only the coalgebra structures. 
As a vector space, the Drinfel'd double $DSU(2)$
 can be identified with the tensor product of the space  $F(SU(2))$ of continuous functions on $SU(2)$ and the group algebra $ \CC(SU(2))$: 
\be
DSU(2) \; \equiv \; D(F(SU(2))) = F(SU(2)) \otimes \mathbb \CC(SU(2)) \, .
\ee
The construction is analogous to the one for a finite group, but the sums over group elements become integrals and Kronecker delta functions  are replaced by distributions.

An alternative definition of the Drinfel'd double of a locally compact group  was developed in \cite{kM,KBM}. In this formulation, the Drinfel'd double $DSU(2)$ 
is  formulated in terms of continuous functions 
 on $SU(2)\times SU(2)$. This formulation was used in \cite{MN1, we2, we3}, as it makes explicit  the close link between the classical and the quantised theory. Its drawback is that the expressions for the Hopf algebra operations become more complicated and the similarities with the Drinfel'd double construction for finite groups are not readily apparent.
 
In this paper, we use both formulations. In the formulation \cite{kM} in terms of functions on $SU(2)\times SU(2)$, the Drinfel'd double $DSU(2)$ is identified with the space of continuous functions $F(SU(2)\times SU(2))$  
and its Hopf-$*$-algebra structure  is given as follows
  \ba
&& \text{Product} \; : \;\label{proda0}
(F_1\cdot F_2)(v,u):=\int_{SU(2)}  \!\!\!\!\!\!dz\; F_1(z,u)\,F_2(z^\inv v,z^{-1}uz)
 \\
&& \text{Coproduct} \; : \; (\Delta F)(v_1, u_1;v_2,u_2)=F(v_1,u_1u_2)\,\delta_{v_1}(v_2) \\
&& \text{Antipode} \; : \; (S F)(v,u)=F(v^\inv, v^{-1}u^{-1}v) \\
&& \text{Unit} \; : \; 1(v,u)=\delta_e(v)\\
&& \text{Counit} \; : \; \varepsilon(F)=\int_{SU(2)} \!\!\!\!\!\!dz\; F(z,e)\\
&& \text{Star structure} \; : \;F^*(v,u)=\overline{F(v^\inv, v^{-1}uv)}\, \label{stara01},
\ea
where   $dz$ denotes the Haar measure on $SU(2)$. The Hopf algebra $DSU(2)$ is a quasi-triangular ribbon Hopf algebra. To describe its universal $R$-matrix and ribbon element, it is necessary to consider not only continuous functions  but also Dirac delta distributions on $SU(2)\times SU(2)$, which can be included by adjoining them \cite{kM,KBM}.   
The universal $R$-matrix then takes the form
\begin{align}
&R(v_1,u_1;v_2,u_2) = \delta_e(v_1)\delta_e(u_1v_2^{-1}).\end{align}
The ribbon elements $c$ is characterised by the ribbon relation $\Delta c= (R_{21} R) \bigl(  c\otimes c \bigr)$, where  $R_{21}=\sigma(R)$ denotes the opposite $R$-matrix. It is given by
\begin{align}
c(v,u) =  \delta_v(u) \,.\label{riba0}
\end{align}
To obtain the formulation in which $DSU(2)$ is identified with the tensor product $F(SU(2))\oo \CC(SU(2))$, one uses the identification
\begin{align}
\label{ident}
\delta_u\oo g\;\;\leftrightarrow\;\; \delta_g\oo \delta_u\qquad\qquad \forall u,g\in SU(2).
\end{align}
Inserting distributions $\delta_u\oo \delta_g$ with $u,g\in SU(2)$  into \eqref{proda0} to \eqref{stara01} then yields the following expressions for the Hopf $*$-algebra structure 
\ba
&& \text{Product} \; : \; (\delta_{u_1} \otimes g_1) (\delta_{u_2} \otimes g_2) \; = \; 
\delta_{g_1u_2g_1^{-1}}\;\delta_{u_1} \otimes g_1g_2 \label{proda2}\\
&& \text{Coproduct} \; : \; \Delta(\delta_u\otimes g) \; = \; 
\int_{SU(2)}  \!\!\!\!\!\!dz\; (\delta_z \otimes g) \otimes (\delta_{z^\inv u}\otimes g)\label{cop} \\
&& \text{Antipode} \; : \; S(\delta_u \otimes g) \; = \; \delta_{g^{-1}u^{-1}g} \otimes g^{-1} \label{ap}\\
&& \text{Unit} \; : \; 1 \otimes e \\
&& \text{Counit} \; : \; \varepsilon(\delta_u \otimes g) \; = \; \delta_e(u) \\
&& \text{Star structure} \; : \; (\delta_u \otimes g)^\star \; = \; 
S(\delta_{u^{-1}} \otimes g) \; = \; \delta_{g^{-1}ug} \otimes g^{-1},
\ea
where $dz$  again denotes the Haar measure on $SU(2)$.
The  universal  $R$-matrix  and the ribbon element  take the form
\ba
R= \int_{SU(2)}  \!\!\!\!\!\!du\;(\delta_u \otimes e) \otimes (1 \otimes u)\qquad\qquad c= \int_{SU(2)}   \!\!\!\!\!\!du\;\delta_u\oo u\label{riba2},
\ea
While some of the expressions \eqref{proda2} to \eqref{riba2} may appear ill-defined due to the presence of products of $\delta$-distributions, they have a clear interpretation in terms of functions in $F(SU(2)\times SU(2))$. 
For this, one expands elements of $DSU(2)$ formally in terms of the singular elements $\delta_u\oo v$
\begin{align}\label{expansion}
F=\int_{SU(2)\times SU(2)} \!\!\!\!\! \!\!\!\!\!\!  \!\!\!\!\!\!dudv\;  F(v,u)\;\delta_u\oo v.
\end{align}
By combining this expression with formulas  \eqref{proda2} to \eqref{riba2} for the quasi-triangular ribbon Hopf algebra structure, one  recovers expressions \eqref{proda0} to \eqref{riba0} as transformation laws of the ``coefficient functions" $F(u,v)$. All expressions involving singular elements of the form $\delta_u\oo v$ are to be interpreted in this sense in the following.

The dual $DSU(2)^*$ of the Drinfel'd double $DSU(2)$ can be identified  with either the
tensor product $DSU(2)^* =  \CC(SU(2)) \oo F(SU(2))$ or the space of functions $F(SU(2)\times SU(2))$. Its Hopf $*$-algebra structure is induced by the one of  $DSU(2)$ via the pairing
\begin{align}\label{pairing}
\langle u\oo \delta_g, \delta_v\oo h\rangle=\delta_v(u)\delta_g(h)\qquad\forall u,v,g,h\in SU(2).
\end{align}
In terms of the singular elements $u\oo \delta_g$ it takes the form
\ba\label{produal}
&& \text{Product} \; : \; ({u_1} \otimes \delta_{g_1}) ({u_2} \otimes \delta_{g_2}) \; = \; 
 u_1u_2\oo \delta_{g_1}\delta_{g_2} \\
&& \text{Coproduct} \; : \; \Delta(u\otimes \delta_g) \; = \; 
\int_{SU(2)}\!\!\!\! dh\; (u\oo \delta_h)\oo (h^\inv u h \oo \delta_{h^\inv g }) \label{codual}\\
&& \text{Antipode} \; : \; S(u \otimes \delta_g) \; = \; g^\inv u^\inv g\oo \delta_{g^\inv}\\
&& \text{Unit} \; : \; e \otimes 1 \\
&& \text{Counit} \; : \; \varepsilon(u \otimes \delta_g) \; = \; \delta_g(e) \\
&& \text{Star structure} \; : \; (u \otimes \delta_g)^\star \; = \; 
u^\inv\oo \delta_g.\label{stardual}
\ea
As in the case of the Drinfel'd double $DSU(2)$, these expressions have a clear interpretation in terms of functions in $F(SU(2)\times SU(2))$. The corresponding identification and expansion in terms of ``coefficient functions"  are given by
 \begin{align}
\label{identdual}
&u\oo \delta_g\;\;\leftrightarrow\;\;\delta_g\oo \delta_u \qquad
&F=\int_{SU(2)\times SU(2)} \!\!\!\!\!\! \!\!\!\!\!\! \!\!\!\!\!\!dudg\; F(g,u)\; u\oo \delta_g.
\end{align}
Using this expansion together with formulas \eqref{produal} to \eqref{stardual}, one obtains expressions for the Hopf $*$-algebra structure in terms of functions in $F(SU(2)\times SU(2))$
\begin{align}
&\text{Product} \; : \;\label{dprod}
(F_1\cdot F_2)(v,u):=\int_{SU(2)} \!\!\!\!dz\; F_1(v,z)\,F_2(v, z^\inv u)
 \\
&\text{Coproduct} \; : \; (\Delta F)(v_1, u_1;v_2,u_2)=F(v_1 v_2,u_1)\,\delta_{u_2}(v_1^\inv u_1 v_1) \\
&\text{Antipode} \; : \; (S F)(v,u)=F(v^\inv, v^{-1}u^{-1}v), \\
&\text{Unit} \; : \; 1(v,u)=\delta_e(u),\\
&\text{Counit} \; : \; \varepsilon(F)=\int_{SU(2)}   \!\!\!\!\!\!dz\;F(e,z)\\
&\text{Star structure} \; : \;F^*(v,u)=\overline{F(v, u^\inv)}\, \label{stara0}.
\end{align}
The Drinfel'd double $DSU(2)$ has two canonical representations, the adjoint representation on itself and the adjoint representation on its dual $DSU(2)^*$. The former is given by 
\begin{align}
\text{ad}(y)(x)=\sum_{(y)} y_{(1)} \cdot x \cdot S(y_{(2)})\qquad \forall x,y\in DSU(2)
\end{align}
where  $\Delta(y)=\sum_{(y)} y_{(1)} \oo y_{(2)}$ denotes the coproduct in Sweedler notation and $S$  the antipode of $DSU(2)$. Inserting expressions \eqref{cop} and \eqref{ap} for the coproduct and antipode yields
\begin{align}\label{ad1}
\text{ad}(\delta_h\oo g)(\delta_u\oo v)=\delta_{g^\inv h g}([u,v^\inv])\;\delta_{gug^\inv}\oo gvg^\inv.
\end{align}
The adjoint representation of $DSU(2)$ on its dual $DSU(2)^*$ is defined by
\begin{align}
\la \text{ad}^*({x}) \alpha, y\ra \; \equiv \; \sum_{(x)} \la \alpha , S(x_{(1)}) \cdot y \cdot x_{(2)}\ra \qquad\forall
x,y\in DSU(2), \alpha \in DSU(2)^*. 
\end{align}
Inserting expressions \eqref{cop} and \eqref{ap} and the paring \eqref{pairing}, one  obtains
\begin{align}\label{ad2}
\text{ad}^*(\delta_v\oo u) (a\oo \delta_b)=\delta_{u^\inv vu}([a^\inv, b^\inv]) \; uau^\inv\oo \delta_{ubu^\inv}.
\end{align}

 \subsubsection{Irreducible representations of  $DSU(2)$}

\label{doublerepsect}

The fact that  $DSU(2)$ is a deformation of  the group algebra $\CC(ISU(2))$ that affects  only the coalgebra structure implies that each  irreducible $*$-representation of $D(SU(2))$  corresponds to  an irreducible representation of the three-dimensional Euclidean group $ISU(2)$.  It  is shown in \cite{kM,KBM} that  the irreducible $*$-representations of $DSU(2)$ are labelled by conjugacy classes $\mathcal C_\mu\subset SU(2)$ and irreducible representations $\pi_s: N_\mu\rightarrow \text{End}(V_s)$ of the centralisers
\begin{align}
\label{centraliser}
N_\mu=\{n\in SU(2)\;|\; ng_\mu n^\inv=g_\mu \},
\end{align}
where $s\in\ZZ/2$ and $g_\mu\in\mathcal C_\mu$ is a fixed element of the conjugacy class $\mathcal C_\mu$.
In the following, we will choose the representatives $g_\mu$ as the diagonal matrices
\begin{align}
g_\mu=e^{\mu J_0}= \left( \begin{array}{cc}e^{i\mu/2} & 0 \\ 0 & e^{-i\mu/2} \end{array}\right)\qquad J_0=\tfrac 1 2 \left(\begin{array}{cc}i  & 0 \\ 0 & -i\end{array}\right).\label{j0def}
\end{align} 
Irreducible $*$-representations of $DSU(2)$ are thus given by two parameters $(\mu,s)$. The continuous parameter $\mu\in[0,2\pi[$ 
defines the $SU(2)$-conjugacy class $C_\mu$ and  is  interpreted as a mass in the context of 3d gravity. The parameter
 $s\in\mathbb Z/2$  labels the irreducible representations of the centraliser $N_\mu$ and is interpreted as an  internal angular momentum or spin.

The representation spaces $V_{\mu,s}$  have the structure of  Hilbert spaces and are given by \cite{kM,KBM}\begin{align}
\label{doublerep0}
 &V_{\mu, s}=\{\psi \in L^2(SU(2), V_s) \vert \psi(g n) = \pi_s(n) \psi(g),\, \forall  n\in N_\mu, \, g\in SU(2)\}\\
 & \langle \psi,\phi \rangle=\int_{SU(2)}  \!\!\!\!du\;\langle\bar\psi(u),\phi(u)\rangle_s,\nonumber
 \end{align}
where $du$ denotes the Haar measure of $SU(2)$ and  $\langle\,,\rangle_s$  the inner product on the representation spaces  $V_s$  of the centralisers. The action of  $D(SU(2))$ on $V_{\mu,s}$ is given by
\begin{align}\label{irrep of DG}
\Pi_{\mu, s} (\delta_u\oo g)\psi(v)=\delta_u(v g_\mu v^\inv)\psi(g^\inv v).\end{align}
 In the generic case  $\mu\notin 2\pi\ZZ$, the centralisers $N_\mu$
 are isomorphic to the group $U(1)$. The Hilbert spaces $V_{\mu,s}$  take the form
 \begin{align}
\label{doublerep}
 &V_{\mu,s}=\{\psi\in L^2(SU(2)) \vert \;\psi(g g_\theta) = e^{is\theta} \psi(g)\;\;\; \forall \theta \in \RR, \,\,\, g\in SU(2)\}\\
& \langle \psi,\phi \rangle=\int_{SU(2)} \bar\psi(u)\phi(u)\;\;du,\nonumber
 \end{align}
and the action of   $DSU(2)$  on $V_{\mu s}$ is given by
 \begin{align}\label{irrep of DG2}
(\Pi_{\mu, s} (\delta_u\oo g)\psi)(v)=\delta_u(vg_\mu v^\inv)\cdot \psi(g^\inv v)\qquad \psi\in V_{\mu s}, \;\;\; u,g\in SU(2).
\end{align}
 For  $\mu\in 2\pi\ZZ$, we have $N_\mu\cong SU(2)$ and the representation spaces $V_{\mu, s}$ become
 \begin{align}
 &V_{0,s}\!=\!\{\psi\in L^2(SU(2), V_s)\,|\, \psi(uv)\!=\!\pi_s(v)\psi(u), \forall u,v\in SU(2)\}\\
 & \langle \psi,\phi\rangle=\langle \psi(e), \phi(e)\rangle_s,\nonumber
 \end{align}
 where $\pi_s=\pi_{-s}: SU(2)\rightarrow \text{End}(V_s)$ is the irreducible unitary $SU(2)$-representation of spin $|s|$. 
The representation spaces $V_{0,s}$ are thus isomorphic to $V_s$, and 
their Hilbert space structure is given by the Hermitian product $\langle\;,\;\rangle_s$ on $V_s$.
 The action of  $D(SU(2))$ on $V_{0,s}$ is given by
 \begin{align}\label{muoact}
\Pi_{0, s}(\delta_u\oo g)\psi(v)=\pi_s(g^\inv v)\psi(e).
\end{align}
It is shown in \cite{kM,KBM}  that for both the generic case $\mu\notin 2\pi\ZZ$ and the singular case $\mu\in 2\pi\ZZ$, a basis of the Hilbert space $V_{\mu s}$ is given by the Wigner functions $D^J_{mn}$ (see appendix \ref{su2char}). The Wigner functions  are directly related to the matrix elements of 
the irreducible $SU(2)$ representations $\rho_J: SU(2)\rightarrow \text{End}(V_J)$, $J\in\NN/2$ 
\begin{align}\label{wigfunc}
D^J_{nm}(g)={\sqrt{ d_J}}\,\langle j,n|\rho_J(g)|m,j \rangle\qquad d_J=\text{dim}(V_J)=2J+1, 
-j\leq m,n\leq j. 
\end{align}
 For the generic case $\mu\notin 2\pi\ZZ$, a Hilbert basis of $V_{\mu s}$ is given by  the Wigner functions  $D^J_{ms}$ with $J\geq |s|$, $-J\leq m\leq J$, on which
 $DSU(2)$  acts according to 
\begin{align}
\Pi_{\mu, s}(\delta_u\oo g) D^J_{ms}(x)=\delta_u(xg_\mu x^\inv)\; D^J_{ms}(g^\inv x)\label{nonsingrep}.
\end{align}
For $\mu\in 2\pi\ZZ$, the Hilbert basis of $V_{\mu,s}$ contains the Wigner functions $D^J_{mn}$ with $J=|s|\in\NN/2$, $-|s|\leq m,n\leq |s|$
and the representation of $DSU(2)$ is given by 
\begin{align}
\Pi_{0,s}(\delta_u\oo g)D^{|s|}_{mn}(x)=\delta_e(u)\;D^{|s|}_{mn}(g^\inv x)\, .\label{mu0act}
\end{align}

\subsection{ Construction of the graph algebra}

\label{combalg}

We have now assembled the relevant structures to construct the graph algebra $\mathcal L(T, ISU(2))$  of the Euclidean torus universe and its  irreducible Hilbert space representations.
For this, we consider a minimal graph on $T$, which is a set of generators of the fundamental group $\pi_1(T)$ as  depicted in Figure \ref{pi1figure}.

The graph algebra  of the torus is generated by two elements $A,B$ associated with the $a$ and $b$-cycle of the torus, which can be interpreted as functions on $SU(2)\times SU(2)$. It can be realised as  a subalgebra
of $F(SU(2)\times SU(2))$ characterised by the reflection relations \cite{AGSI,AGSII,AS,BR}. In order to give a concise description of these equations and to exhibit the underlying mathematical structures, it is convenient to  characterise it as a subalgebra of the tensor product $\mathcal L(T,ISU(2))\oo DSU(2)$. Using the expansion \eqref{expansion} of elements of $DSU(2)$ in terms of the singular elements $\delta_u\oo v$, we set
\begin{align}\label{abexp}
A=\int_{SU(2)\times SU(2)}\!\!\!\!\!\!\!\!\!\!\!\!\!\!\!dudv\; A(v,u)\; \delta_u\oo v\qquad B=\int_{SU(2)\times SU(2)}\!\!\!\!\!\!\!\!\!\!\!\!\!\!\!dudv\; B(v,u)\; \delta_u\oo v
\end{align}
where $A(v,u)$ and $B(v,u)$ are functions on $SU(2)\times SU(2)$ that correspond to elements of the graph algebra $\mathcal L(T,ISU(2))$.
The formulas for the reflection relations from 
 \cite{AGSI,AGSII,AS,BR} read
\ba\label{refl_aa}
A_1 \, R_{21} \, A_2 \, R_{21}^{-1} & = & R_{12}^{-1}\, A_2 \, R_{12} \, A_1 \,,\\
B_1 \, R_{21} \, B_2 \, R_{21}^{-1} & = & R_{12}^{-1}\, B_2 \, R_{12} \, B_1 \,,\label{refl_bb}\\
R_{12}\, A_1\, R_{21}\, B_2 & = & B_2 \, R_{12} \, A_1 \, R_{12}^{-1} \label{refl_ab}\;,
\ea
where $A_1=A\oo 1$, $B_1=B\oo 1$, $A_2=1\oo A$, $B_2=1\oo B \in DSU(2)\times DSU(2)$,
  $R$ is the universal $R$-matrix 
of $DSU(2)$ and $R_{21}$ its opposite. By inserting 
the  expansion  \eqref{abexp} into \eqref{refl_aa} to \eqref{refl_ab}, we obtain a reformulation of the reflection relations  in terms of functions on $SU(2)\times SU(2)$
\ba\label{refl_aa2}
&& A(v_1[v_2^{-1},u_2],u_1) \, A(v_2,u_2) \; = \; A(u_1v_2u_1^{-1},u_1u_2u_1^{-1}) \, A(v_1,u_1)\;,\\
&& B(v_1[v_2^{-1},u_2],u_1) \, B(v_2,u_2) \; = \; B(u_1v_2u_1^{-1},u_1u_2u_1^{-1}), \, B(v_1,u_1)\label{refl_bb2}\;,\\
&& A(v_1u_1^{-1}u_2^{-1}u_1,u_1) \, B(u_1^{-1}v_2,u_1^{-1}u_2u_1) \; = \; B(v_2[v_1^{-1},u_1],u_2) \, A(u_1,v_1) \label{refl_ab2}.
\ea
The graph algebra  of the torus can then be identified with the associative algebra that is generated multiplicatively by functions $A,B$ on $SU(2)\times SU(2)$ subject to the reflection equations \eqref{refl_aa2} to \eqref{refl_ab2}.
In this description the graph algebra  appears rather complicated and it is
not readily apparent  how to classify its irreducible Hilbert space representations.

The central result that allows one to address this issue is Alekseev's theorem \cite{Alekseev} stating that the graph algebra $\mathcal L(\Sigma,G)$ associated with a general $n$-punctured genus $g$ surface $\Sigma$ is isomorphic
to the tensor product $H(\mathcal A)^{\oo g}\oo \mathcal A^{n}$, where $\mathcal A$ is the quantum group associated with the gauge group $G$ and $H(\mathcal A)$ its Heisenberg double algebra.  In the case at hand, we have $g=1$, $n=0$, and the relevant quantum group is the Drinfel'd double $DSU(2)$. The Alekseev isomorphism thus maps the graph algebra  to the Heisenberg double $H(DSU(2))$.
The general definition of the Heisenberg double of a Hopf algebra $\mathcal A$ is given in Def.~\ref{heisdef} in Appendix \ref{hopfapp}. By specialising this definition to the Hopf algebra  $\mathcal A=DSU(2)$ we obtain
\begin{definition}\label{hdsu2}
The Heisenberg double $H(DSU(2))$ is an associative unital algebra  that as a vector space is isomorphic to the tensor product $DSU(2)\oo DSU(2)^*$. The Hopf algebras $DSU(2)$ and $DSU(2)^*$ can be embedded into $H(DSU(2))$ by injective algebra homomorphisms. 
In terms of the singular elements $\delta_u\oo v\in DSU(2)\subset H(DSU(2))$ and $u\oo \delta_v\in DSU(2)^*\subset H(DSU(2))$, its multiplicative structure is given by 
\begin{align}\label{hmult}
&(\delta_{u_1} \otimes g_1)\cdot (\delta_{u_2} \otimes g_2) = (\delta_{u_1}\delta_{g_1u_2g_1^{-1}} \otimes g_1g_2) \\
&(v_1\otimes \delta_{h_1}) \cdot (v_2\otimes \delta_{h_2}) =  v_1v_2 \otimes \delta_{h_1}\delta_{h_2} \nonumber\\
&(v\otimes \delta_h)\cdot (\delta_u\otimes g) =  (\delta_{h^{-1}vhu} \otimes g)\oo (v\otimes \delta_{hg}),\nonumber
\end{align}
where we use the shorthand notation
\ba
(\delta_u\otimes g) & \equiv & (\delta_u \otimes g) \otimes (e \otimes 1) \in DSU(2) \subset H(DSU(2)) \\
(v\otimes \delta_h) & \equiv & (1 \otimes e) \otimes (v \otimes \delta_h) \in DSU(2)^* \subset H(DSU(2)).\nonumber
\ea
The inclusions of $DSU(2)$ and $DSU(2)^*$ into $H(DSU(2))$ define a star structure on $H(DSU(2))$, which is given by 
\begin{align}\label{hdstar}
\left((\delta_u\oo g)\oo(v\oo \delta_h)\right)^*=(\delta_{h^\inv v^\inv h g^\inv u g}\oo g^\inv)\oo (v^\inv\oo \delta_{hg^\inv ug}).
\end{align}
\end{definition}
Using this definition of the Heisenberg double, we can determine an explicit expression for the Alekseev isomorphism for the Drinfel'd double $DSU(2)$, which is given in the following lemma.
\begin{lemma}\label{alisom}
The Alekseev isomorphism $\Gamma:{\mathcal L}(T, ISU(2))\rightarrow H(DSU(2))$ between the graph algebra $\mathcal L(T, ISU(2))$ of the torus and the Heisenberg double $H(DSU(2))$ is given by
\begin{align}\label{isoL_HDG}
\Gamma:\qquad
&A(v,u)  \mapsto \int_{SU(2)}\!\!\! dw \, (\delta_{uwvu^{-1}}\otimes u)\otimes (u\otimes \delta_{uw^{-1}}) \;\;\; \quad B(v,u)   \mapsto  \delta_{uvu^{-1}} \otimes u.
\end{align}
\end{lemma}

{\bf Proof:} To prove that \eqref{isoL_HDG} defines an algebra isomorphism, we need
to show that the images $\Gamma(A(v,u))$, $\Gamma(B(v,u))$ satisfy the reflection equations \eqref{refl_aa2} to \eqref{refl_ab2} and form a basis of $H(DSU(2))$. The former can be demonstrated by direct calculation using \eqref{isoL_HDG} and the multiplication law \eqref{hmult} of $H(DSU(2))$. To show that they form a basis of $H(DSU(2))$, we express the basis elements $(\delta_v\oo u)\oo (e\oo 1)$ and $(1\oo e)\oo(u\oo \delta_v)$ in terms of the images $\Gamma(A(v,u))$, $\Gamma(B(v,u))$. This yields
\begin{align}
&(\delta_v\oo u)\oo (e\oo 1)=\Gamma(B(u^\inv v u,u))\\
&(1\oo e)\oo (u\oo \delta_v)=\int_{SU(2)}\!\!\!\!\!dz\; \Gamma(B(z,u^\inv))\Gamma(A(u^\inv v u^\inv z u, u)).\nonumber\qquad\qquad\qquad\qquad\qquad\qquad\Box
\end{align}

\subsection{Representations of the graph algebra - kinematical states}
\label{kinstates}

The Alekseev isomorphism between the graph algebra of the torus and the Heisenberg double $H(DSU(2))$ allows one to directly construct the  irreducible representations of the graph algebra. For this, one makes use of the representation theory of the Heisenberg double algebras which was  first investigated in \cite{AF} and is summarised in Theorem.~\ref{rephdth}. 
By adapting these  results to the quantum group $DSU(2)$, one obtains the following lemma.
\begin{lemma}
The Heisenberg double $H(DSU(2))$ has a single irreducible Hilbert space representation, which is realised on the dual $DSU(2)^*$ equipped with  the inner product\begin{align}\label{innprod}
\langle F,G \rangle=\int_{SU(2)\times SU(2)}\!\!\!\!\!\!\!\!\!\!\!\!\!\!\!dudv\; \overline{F(v,u)}G(v,u)\qquad\qquad\langle u\oo \delta_v, g\oo \delta_h\rangle=\delta_u(g)\delta_v(h).
\end{align}
In terms of the singular elements $(u\oo \delta_g)\equiv(1\oo e)\oo(u\oo\delta_g)$, $(\delta_u\oo g)\equiv(\delta_u\oo v)\oo (e\oo 1)\in H(DSU(2))$ and $v\oo \delta_h\in DSU(2)^*$  this representation is given by
\begin{align}
\label{hdrepdouble}
\pi(u\oo \delta_g)(v\oo \delta_h)=uv\oo \delta_g\delta_h\qquad \pi(\delta_u\oo g)v\oo \delta_h =\delta_u(gh^\inv vhg^\inv) v\oo \delta_{hg^\inv}.
\end{align}
The irreducible representation \eqref{hdrepdouble} is a $*$-representation  with respect to the inner product \eqref{innprod} and the star structure \eqref{hdstar}:
\begin{align}
\langle \pi(G^*)\, a\oo \delta_b, c\oo \delta_d\rangle=\langle a\oo \delta_b, \pi(G)\,c\oo \delta_d\rangle\qquad\forall G\in H(DSU(2)).\label{starrep}
\end{align}
\end{lemma}
Together with the Alekseev isomorphism \eqref{isoL_HDG}, this lemma allows one to classify all irreducible $*$-representations of the graph algebra.
\begin{corollary}\label{graphreps}
The graph algebra $\mathcal L(T, ISU(2))$ has a single irreducible $*$-representation. The representation space  is the dual $DSU(2)^*$ equipped with the canonical inner product \eqref{innprod}. The action of its generators $A,B$ in this representation is given by
\begin{align}\label{graphrep}
&\pi(A(v,u)) (a \otimes \delta_b)   =    \delta_v(ab) \, (ua \otimes \delta_{bu^{-1}}) \quad
\pi(B(v,u)) (a \otimes \delta_b)  = \delta_v(b^{-1}ab) \, (a \otimes \delta_{bu^{-1}}).
\end{align}
It is a  $*$-representation with respect to the inner product \eqref{innprod} and  the star structure
\begin{align}\label{stargraph}
X(v,u)^*=X(uvu^\inv, u^\inv)\qquad X=A,B.
\end{align}
\end{corollary}
We have thus constructed the graph algebra  which is the quantum counterpart of Fock and Rosly's Poisson algebra given by \eqref{poissj} to \eqref{poissfj}.
 It serves as the starting point for the construction of the moduli algebra $\mathcal M(T, ISU(2))$ of gauge invariant quantum observables. Corollary \eqref{graphreps} defines the kinematical (non-gauge invariant) Hilbert space of the theory.


\section{Implementation of the constraints}
\label{physhilb}

\subsection{The quantum flatness constraint}
\label{constsect}
In combinatorial quantisation, the physical (gauge invariant) Hilbert space is constructed via Dirac's constraint quantisation formalism by promoting the flatness constraint \eqref{cdef} to a constraint operator. This operator is an element of the graph algebra and acts on the kinematical Hilbert space  in Corollary \ref{graphreps}. 
The physical (gauge invariant) Hilbert space is then to be identified with the linear subspace spanned by the states that are invariant under the action of this constraint operator.

The first step in  this construction is the definition of the quantum flatness constraint.  Following the formalism  in \cite{AGSI,AGSII, BNR}, we define this constraint operator as 
\begin{align}\label{combflat}
C \; \equiv \; \nu^2  AB^{-1}A^{-1}B \; = \; 1\quad\quad\nu^2=c^\inv S(c^\inv)=\int_{SU(2)} \!\!\!\!\!\!du\; \delta_u\oo u^{-2}
\end{align}
where $A$ and $B$ are viewed as elements of ${\cal L}\otimes DSU(2)$ defined as in \eqref{abexp} and $\nu$ is given by the ribbon element. 
To obtain an explicit expression in terms of functions $A(v,u)$, $B(v,u)$ on $SU(2)\times SU(2)$ we expand the elements $A,B$ in terms of singular elements $\delta_u\oo v$ according to \eqref{abexp} and use the multiplication law \eqref{hmult}. This yields
\begin{align}
C(v,u )\!\!=\!\!\!\!\int_{SU(2)^3}\!\!\!\!\!\!\!\!\!\!\!\!\!dv_1dv_2dv_3\,A(v_1,u)B^{-1}(v_1^{-1}v_2,v_1^{-1}uv_1)     A^{-1}(v_2^{-1}v_3,v_2^{-1}uv_2)B(v_3^{-1}u^2v,v_3^{-1}uv_3).\!\label{cexpr}
    \end{align}
To evaluate this expression further, we determine
 the inverses of the elements $A(v,u)$, $B(v,u)\in H(DSU(2))$. This is most easily done via Alekseev isomorphism   \eqref{isoL_HDG} which relates the graph algebra to the Heisenberg double  $H(DSU(2))$. This yields
\begin{align}
A^{-1}\!=&\!\!\int_{SU(2)\times SU(2)}\!\!\!\!\!\!\!\!\!\!\!\!\!\!\!\!\!\!\!\!\!\!\!\!\! dudv\,  A^{-1}(v,u) \delta_u \otimes v \qquad A^{-1}(v,u)\!= \!\!\!\int_{SU(2)} \!\!\!\!\! \!\!\!\!\! dh\;(\delta_{hv^{-1}} \!\otimes\! v^{-1}u^{-1}v) 
\!\otimes\! (v^{-1}u^{-1}v \!\otimes\! \delta_{h^{-1}v^{-1}u^{-1}v})\nonumber\\
B^{-1}\!=&\!\!\int_{SU(2)\times SU(2)}\!\!\!\!\!\!\!\!\!\! \!\!\!\!\!\!\!\!\!\!\!\!\!\!\!\!dudv\; B^{-1}(v,u) \delta_u \otimes v\qquad B^{-1}(v,u) \!=\  \delta_{v^{-1}} \otimes v^{-1}u^{-1}v. \label{abinv}
\end{align}
Inserting these expressions into \eqref{cexpr}  yields an explicit expression for  the quantum flatness constraint
 \begin{align}\label{constraintC}
  C(v,u)=&\int_{SU(2)\times SU(2)}\!\!\!\!\! \!\!\!\!\!\!\!\!\!\!\!\!\!\!\!\! da db\; \delta_{uvu^\inv}([a,b]) 
    (\delta_{a^\inv uvu^\inv}\otimes [b,u^{-1}])\otimes([a^\inv,u^\inv] \otimes \delta_{u^{-1}b^\inv u}).
\end{align}
This  allows us to determine the action of the constraint $C(v,u)$ on the kinematical Hilbert space $DSU(2)^*$.  Using expression \eqref{hdrepdouble} for the irreducible representation of the Heisenberg double $H(DSU(2))$ we obtain
\begin{align}\label{constexp}
\pi(C(v,u))(a\oo \delta_b)=\delta_v([a^\inv, b^\inv])\; uau^\inv\oo \delta_{ubu^\inv}.
\end{align}
 Imposing invariance under the action of the constraint $C(v,u)$ amounts to setting
\begin{align}\label{ccond}
\pi(C(v,u))\psi=\epsilon(\delta_v\oo u)\psi=\delta_e(v)\psi\quad \psi\in DSU(2)^*.
\end{align}
The advantage of the combinatorial  quantisation formalism is that it  relates the implementation of the flatness constraint to the representation theory of the Drinfel'd double $DSU(2)$. More specifically, the
 invariance condition \eqref{ccond} can be reformulated in terms of the adjoint representation of $DSU(2)$ on its dual $DSU(2)^*$. By comparing  the action of the quantum flatness constraint in \eqref{constexp} to expression \eqref{ad2} for the adjoint action of $DSU(2)$ on $DSU(2)^*$, we find that the two agree up to a replacement $v\mapsto u v  u^\inv$.
Condition \eqref{ccond} which encodes the invariance of  states $\psi\in DSU(2)^*$  under the
action of the constraint operator  thus amounts  to imposing invariance under the adjoint action \eqref{ad2} of $DSU(2)$ on $DSU(2)^*$.

Moreover, one finds that the subalgebra $\cal C$ of the graph algebra that is  generated multiplicatively by the elements $C(v,u)$ is isomorphic to the Drinfel'd double algebra $DSU(2)$. An isomorphism between the two algebras is given by
\begin{eqnarray}
\Upsilon \; : \; {\mathcal C} \; \longrightarrow \; DSU(2) \;\;\;\;\;\;\;
C(v,u) \, \longmapsto \, \delta_{uvu^\inv} \otimes u \,.
\end{eqnarray}
The combinatorial quantisation formalism thus allows one to reduce the implementation of the flatness constraint  to the construction of the subspace $\mathcal H_{p}\subset DSU(2)^*$
 which is invariant under the adjoint action of $DSU(2)$ on $DSU(2)^*$.
  
\subsection{Gauge invariant states}  \label{combstates}
  
The fact that invariance under the action of the constraint operator is equivalent to invariance under the adjoint action  of $DSU(2)$ on $DSU(2)^*$ allows one to directly construct the gauge invariant Hilbert space. The central idea is to use the characters $\chi_{\mu, s}$ of the irreducible representations of $DSU(2)$ to construct a Hilbert basis. 
The characters have a natural interpretation as elements of the dual $DSU(2)^*$  and are invariant under the adjoint action. They were first
 derived in \cite{kM, KBM} for the description  
 of $DSU(2)$ in terms of functions on $SU(2)\times SU(2)$. 
 In the notation introduced in Sect.~\ref{qugroup} they  take the form 
\begin{align}\label{funcchar}
 \chi_{\mu,s}(F)=\int_{SU(2)} \!\!\!\!dz \int_{N_\mu}\!\!\!\!dn \; F( znz^\inv, z g_\mu z^\inv)\chi^s_{N_\mu}(n),
\end{align}
 where $dn$ denotes the Haar measure of the centraliser $N_\mu$ and $\chi_{N_\mu}^s$ the characters of $N_\mu$ in its irreducible representation labelled by  $s$. The characters of the  irreducible representations of $DSU(2)$ thus can be identified with distributions on $SU(2)\times SU(2)$. As such, they have a natural interpretation as elements of the dual $DSU(2)^*$. 
 The corresponding expressions are obtained by using the identification \eqref{identdual} and by inserting singular elements of the form $\delta_u\oo \delta_g$ into \eqref{funcchar}\begin{align}\label{chardef}
\chi_{\mu,s}=\int_{SU(2)\times SU(2)} \!\!\!\!\!\!\!\!dudg \;\chi_{\mu,s}(\delta_u \oo\delta_g)\;\;u\oo\delta_g\;=\int_{SU(2)}\!\!\!\!\!\!dz\int_{N_\mu}\!\!\!dn\; \chi^s_{N_\mu}(n)\; zg_\mu z^\inv\oo \delta_{znz^\inv}.
\end{align}
In the generic case of masses $\mu\notin 2\pi\ZZ$, for which $N_\mu\cong U(1)$, this expression reduces to
\begin{align}\label{genchar}
\chi_{\mu,s}=\int_{SU(2)} \!\!\!dz\int_{0}^{2\pi} d\theta \;e^{is\theta} \;zg_\mu z^\inv\oo \delta_{ zg_\theta z^\inv},
\end{align}
where $g_\mu,g_\theta$ are defined by \eqref{j0def}. The corresponding expression for  $\mu\in 2\pi\ZZ$ is given by
\begin{align}
\chi_{0,s}=\int_{SU(2)\times SU(2)} \!\!\!\!\! dndz\;\; \chi_s(n) e\oo \delta_{znz^\inv} \; = \; 
\int_{SU(2)}  \!\!\! dn \;\; \chi_s(n) \, e \oo \delta_n \, = \, e \oo \chi_s \,,
\end{align}
where $\chi_s$ denotes the characters of  the irreducible $SU(2)$-representation $\pi_{s}=\pi_{-s}: SU(2)\rightarrow \text{End}(V_s)$ of spin $|s|$.
Using the left-invariance of the Haar measure on $SU(2)$, one can show that the characters are invariant under the  adjoint action  \eqref{ad2}  of $DSU(2)$ on $DSU(2)^*$
\begin{align}\label{charcond}
\ad^*(\delta_u\oo g)\chi_{\mu, s}=\delta_u(e)\; \chi_{\mu, s}. 
\end{align}
The characters  thus define  gauge invariant states and serve as building blocks in the construction of the gauge invariant Hilbert space $\mathcal H_{inv}$.  In particular, they define an inner product on $\mathcal H_{inv}$ via 
\begin{align}\label{charinn}
\langle \chi_{\mu,s}, \chi_{\mu',s'}\rangle=\int_{SU(2)\times SU(2)}\!\!\!\!\!\!\!\!\!\!\!\!dudv\;\bar \chi_{\mu, s}(\delta_u \otimes \delta_v)
\chi_{\mu',s'}(\delta_u \oo \delta_v)=\delta_\mu(\mu')\delta_{s,s'},
\end{align}
where $\delta_{s,s'}$ denotes the Kronecker delta and $\delta_\mu$ the $2\pi$-periodic  delta distribution on $\RR$. 

In the following, it will be advantageous to work with an alternative characterisation of the gauge invariant  Hilbert space  in terms of states which depend on two continuous parameters $\alpha,\beta$. These states  can be viewed as Fourier transforms of the characters in  \eqref{genchar} and are given by
 \begin{align}\label{chimunudef}
 \chi_{\alpha,\beta}= \int_{SU(2)}\!\!\! \!\!\!\!dz \;\;
zg_\alpha z^\inv \oo \delta_{zg_\beta z^\inv},
 \end{align}
Formula \eqref{genchar} provides an explicit expression for the  characters $\chi_{\mu,s}$ in terms of the states $\chi_{\alpha,\beta}$. Moreover, it follows from the 
completeness relation for the characters $\chi_{N_\mu}^s$ of the centralisers  that the states  $\chi_{\alpha,\beta}$ in \eqref{chimunudef} are given in terms of  the states 
 $\chi_{\mu,s}$  as
 \begin{align}\label{FTbasis}
\chi_{\alpha, \beta} = \frac{1}{2\pi} \, \sum_{s\in\ZZ/2} \bar \chi_{N_\alpha}^s(g_\beta) \chi_{\alpha, s}.
\end{align}
The states $\chi_{\mu,\nu}$ thus provide an equivalent characterisation of the gauge invariant Hilbert space  and can be interpreted as Fourier transforms of the characters $\chi_{\mu,s}$. 
 Using expression \eqref{FTbasis}, the completeness relation for the characters $\chi_{N_\mu}^s$ and formula \eqref{charinn} for the inner product, one finds that their inner product is given by
\begin{align}\label{char2}
\langle \chi_{\alpha,\beta},\chi_{\gamma,\delta}\rangle\!=\!\!\!\!\!\!\sum_{s,t\in\ZZ/2} \!\!\!\bar \chi_{N_\alpha}^s(g_\beta)\chi_{N_\gamma}^t(g_\delta)\langle\chi_{\alpha,s},\chi_{\gamma,t}\rangle\!=\!\delta_\alpha(\gamma)\!\!\!\sum_{s\in\ZZ/2} \!\!\bar\chi_{N_\alpha}(g_\beta)\chi_{N_\alpha}^t(g_\delta)\!=\!\delta_{\alpha}(\gamma)\delta_\beta(\delta)
\end{align}
Although expressions \eqref{char2}, \eqref{charinn}  appear ill-defined at first sight due to the terms involving Dirac delta distributions, they can be given a precise meaning. For this, we expand gauge invariant states formally in terms of  the states $\chi_{\alpha,\beta}$ and ``coefficient functions" on $\RR^2$ which are $2\pi$-periodic in both arguments
\begin{align}\label{FTstates}
\psi = \int_0^{2\pi} \!\!\!d\alpha \int_0^{2\pi} \!\!\!d\beta \; \psi(\alpha+\beta, \alpha) \, \chi_{\alpha,\beta}.
\end{align}
The inner product \eqref{char2} then induces an inner product on the space of coefficient functions, and we obtain the following definition of the gauge invariant Hilbert space.

\begin{definition}\label{hinvth}
The gauge invariant Hilbert space  of the Euclidean torus universe is the space of $L^2$-functions on the torus
\begin{align}
&\mathcal H_{inv}=\{\psi: \RR^2\rightarrow \CC\;|\; \psi(\alpha+2\pi m, \beta+2\pi n)=\psi(\alpha,\beta)\;\forall m,n\in\ZZ\}/\sim\\
&\psi\sim\phi\quad\text{if}\quad \langle\psi-\phi,\psi-\phi\rangle=0\nonumber,
\end{align}
 equipped with the standard inner product
\begin{align}\label{FTinnprod}
\langle \phi,\psi\rangle=\int_0^{2\pi}\!\!d\alpha\int_0^{2\pi}\!\!d\beta\; \overline{ \phi(\alpha,\beta)}\psi(\alpha,\beta).
\end{align}
\end{definition}

\subsection{Gauge invariant observables}\label{combob}

We are now ready to construct the gauge invariant observables that  act on the gauge invariant  Hilbert space $\mathcal H_{inv}$. By definition, these are the elements of the graph algebra which commute with the flatness constraint \eqref{constraintC}.
More precisely, 
the algebra $\cal C$ generated by the constraint elements  $C(g,h)$ in \eqref{constraintC} defines an automorphism of the graph algebra 
\begin{align}
{\mathcal C} \times {\mathcal L}(T,ISU(2))  \rightarrow {\mathcal L}(T,ISU(2)) \qquad \, (C;X)\mapsto C\cdot X\cdot C^\inv,
\end{align}
where the multiplication $\cdot$ is  given by \eqref{hmult}, \eqref{isoL_HDG} and \eqref{constraintC}.
It follows from the discussion in Sect.~\ref{constsect} that invariance under this automorphism is equivalent to invariance 
under the following action of $DSU(2)$ on ${\cal L} \oo DSU(2)$
\begin{align}\label{grautom}
DSU(2)\times({\mathcal L}(T,ISU(2))\oo DSU(2)) \; &\rightarrow  {\mathcal L}(T,ISU(2)) \oo DSU(2)\\
(\delta_h\oo g;X) &\mapsto \int_{SU(2)\;\;}\!\!\!\!\!\!dudv\; X(v,u)\;\;\text{ad}(\delta_h\oo g)\delta_u\oo v.\nonumber
\end{align}
By inserting expression
 \eqref{ad1} for the adjoint action of $DSU(2)$ on itself into \eqref{grautom}, we obtain an explicit expression for the action of $DSU(2)$ on the graph algebra
 \begin{align} \label{adac}
(\delta_h\oo g \,, \, X(v,u)) \mapsto \delta_{g^\inv h g}([u,v^\inv])\; X(g^\inv v g,g^\inv u g) \,.
\end{align}
A generic set of observables that are invariant under this action are the Wilson loop  observables associated with elements of the graph algebra. In the combinatorial quantisation formalism for compact gauge groups they are constructed from the characters of the irreducible representations  of the associated $q$-deformed universal enveloping algebras. In the generalisation to the non-compact case, the corresponding expressions require a regularisation. Roughly speaking, matrix elements of the universal enveloping algebras are replaced by functions, Kronecker delta functions by distributions and sums over discrete parameters labelling irreducible representations by integrals. The latter requires the choice of a suitable measure on $SU(2)\times SU(2)$ which is to be determined from structural requirements on the resulting quantum observables.

By generalising the definitions of the combinatorial quantisation formalism  in this way, we obtain the following definition of the Wilson loop observables
\begin{align}\label{wdef0}
W_{\mu, s}(X)=\!\!\int_{SU(2)\times SU(2)}\!\!\!\!\!\!\!\!\!\!\!\!\!\!\!\!\!\!\!\!\!\!dudv\; M(u,v) \, \chi_{\mu,s}(\delta_u \oo \delta_v) X(v,u)
\end{align}
where $X$ is an element of the graph algebra and $M: SU(2)\times SU(2)\rightarrow \RR$ is a  function which satisfies the invariance property 
$M(u,v)=M(g^\inv u g,g^\inv v g)$ $\forall g \in SU(2)$ and defines a measure  on $SU(2)\times SU(2)$. 
Inserting expression \eqref{chardef}  for the characters into \eqref{wdef0}, one  finds that the Wilson loop observables are given by
\begin{align}\label{wdef}
W_{\mu, s}(X) =\!\!\int_{SU(2)}\!\!\!\!\!dz\int_{N_\mu} \!\!\!\!dn\;\; M(g_\mu,n) \, \chi_{N_\mu}^s(n)\, X(znz^\inv, z g_{\mu} z^\inv).
\end{align}
We will now demonstrate that the ambiguity resulting from the choice of the measure $M$  is resolved  by  requiring  that the representation of the Wilson loop observables $W_{\mu,s}(A), W_{\mu,s}(B)$ on the gauge invariant Hilbert space $\mathcal H_{inv}$ is a unitary *-algebra representation.

For this, we need to determine the action of the Wilson loop observables $W_{\mu,s}(A), W_{\mu,s}(B)$ on  $\mathcal H_{inv}$.
By combining \eqref{wdef} and formula \eqref{graphrep} for the representation of the variables $A(v,u), B(v,u)$ on $DSU(2)^*$ we obtain after some computations
 \begin{align}
\pi(W_{\mu,s}(A))\chi_{\alpha,\beta}
=&\,w_{\mu,s}(\alpha+\beta)\,  \chi_{\alpha+\mu,\beta-\mu}\label{aact}\qquad \pi(W_{\mu,s}(B))\chi_{\alpha,\beta}\!=w_{\mu,s}(\alpha)\cdot \chi_{\alpha,\beta-\mu}
\end{align}
where $w_{\mu,s}: \RR\rightarrow \RR$ is the $2\pi$-periodic function given by 
\begin{align}\label{factdef}
w_{\mu,s}(\gamma)= \int_{SU(2)} \!\!\!\!dy\int_{N_\mu} \!\!\!\!dn\; M(g_\mu,n) \chi^s_{N_\mu}(n)\delta_{yny^\inv}(g_{\gamma}).
\end{align}
We will now show that the requirement that the representations given by \eqref{aact} are unitary $*$-representations determine the function $w_{\mu,s}$ uniquely. For this we note that
the associative, unital  algebra $\mathcal A_{inv}$ generated by the  Wilson loop observables  $W_{\mu,s}(A)$, $W_{\mu,s}(B)$  
inherits a canonical  star structure from the star structure \eqref{stardual} on $DSU(2)^*$. It takes the form
\begin{align}\label{obsstar}
&W_{\mu,s}(X)^\star =W_{2\pi-\mu, -s}(X)
\qquad X=A,B.
\end{align}
The requirement that the representation given by \eqref{aact} is a $*$-representation with respect to this star structure and inner product \eqref{char2} reads 
\begin{align}\label{star1}
\langle \pi(W_{\mu,s}(X)^*)\chi_{\alpha,\beta}\,,\, \chi_{\gamma,\delta}\rangle=\langle \chi_{\alpha,\beta}\,,\,\pi(W_{\mu,s}(X))\chi_{\gamma,\delta}\rangle\qquad X=A,B.
\end{align}
The condition that it is a unitary $*$-representation yields the additional requirement
\begin{align}\label{un1}
\pi(W_{\mu,s}(X^*))\pi(W_{\mu,s}(X))\,\chi_{\alpha,\beta}=\chi_{\alpha,\beta}\qquad X=A,B.
\end{align}
By combining these conditions and comparing them with expressions  \eqref{aact} and the definition \eqref{factdef}, we obtain the following lemma.

\begin{lemma} \label{unlemma} The moduli algebra $\mathcal A_{inv}$   generated by the Wilson loop observables $W_{\mu,s}(A)$, $W_{\mu,s}(B)$ acts
on the gauge invariant Hilbert space $\mathcal H_{inv}$ via \eqref{aact}. 
This action defines  a unitary $*$-representation with respect to the inner product \eqref{char2} and the star structure \eqref{obsstar} if and only if
\begin{align}
w_{\mu,s}(\gamma)=\begin{cases} e^{is\gamma} & \mu\notin 2\pi\ZZ\\ 1 & \mu\in 2\pi\ZZ\end{cases}\qquad\forall \mu,\gamma\in[0,2\pi], s\in\ZZ/2.
\end{align}
\end{lemma}
{\bf Proof:}
A short calculation shows that conditions \eqref{star1}, \eqref{un1} are equivalent to the following conditions on the functions $w_{\mu,s}$
\begin{align}\label{starcond}
&\overline{ w}_{2\pi-\mu,-s}(\gamma)=w_{\mu,s}(\gamma)\qquad\qquad
w_{\mu,s}(\gamma)\cdot w_{2\pi-\mu,-s}(\gamma)=1.
\end{align}
To determine how conditions \eqref{starcond} restrict the measure $M(u,v)$ and fix the constants $w_{\mu,s}(\gamma)$, we have to evaluate expression \eqref{factdef}.
For the singular case $\mu\in 2\pi\ZZ$, this can be done by expressing the class function given by the integral over the delta-distribution in \eqref{factdef} in terms of the characters $\chi_s$ of the irreducible representations of $SU(2)$
\begin{align}
f(n) =\int_{SU(2)} \!\!\!\! dy \; \delta_{y n y^\inv}(g_{\alpha}) =\sum_{t\in\NN_0/2} f_t(g_\alpha)\chi_t(n).
\end{align}
Inserting this expression into \eqref{factdef} yields
\begin{align}
w_{\mu,s}(\alpha)=\sum_{t\in\NN_0/2} \int_{SU(2)}\!\!\!\!dn\; M(e, n)\chi_s(n) \chi_t(n) f_t(g_\alpha)=M(e, g_\alpha)f_s(g_\alpha),
\end{align}
where $f_s(\alpha), M(e, g_\alpha)\in\RR$, the characters $\chi_s$ and coefficients $f_s$ are extended to negative spins $s$ via $\chi_s=\chi_{-s}$, $f_{-s}=f_s$, and we used the orthogonality relation \eqref{orthoirr} for the characters of $SU(2)$. The coefficients $w_{\mu,s}$ thus satisfy the first condition in \eqref{starcond}. The second condition in \eqref{starcond}, which encodes the unitarity of the representation, then implies $w_{\mu,s}(\alpha)=1$.

To determine the factor $w_{\mu,s}(\alpha)$ in the generic case $\mu\notin 2\pi\ZZ$, we note that
\begin{align}\label{hep}
\int_{SU(2)} \!\!\!\! dy \; \delta_{yg_\theta y^\inv}(g_{\alpha}) \; = \; N(\alpha) \; \delta_\alpha(\theta)
\end{align}
with a $2\pi$-periodic function $N: \RR\rightarrow \RR$. Inserting this  into the expression  for $w_{\mu,s}(\alpha)$   yields
\begin{align}
w_{\mu,s}(\alpha)=M(g_\mu, g_\alpha)N(\alpha) e^{is\alpha}.
\end{align}
As $M(g_\mu, g_\alpha)N(\alpha)\in \RR$, the first condition in \eqref{starcond} is satisfied if and only if $M(g_\mu, g_\alpha)=M(g_{-\mu}, g_\alpha)$. The
 unitarity condition in \eqref{starcond} then  implies $w_{\mu,s}(\alpha)=e^{is\alpha}$.\hfill $\Box$

By combining the results for the generic and the singular case and inserting the resulting expressions for the measure into \eqref{aact}, we obtain explicit expressions for the action of the gauge invariant observables $W_{\mu,s}(A), W_{\mu,s}(B)$ on the 
states $\chi_{\alpha,\beta}$ which span the gauge invariant Hilbert space $\mathcal H_{inv}$. After reformulating this result in terms of the ``coefficient functions"  in \eqref{FTstates} in terms of functions in $\mathcal H_{inv}$ we obtain the following theorem.
 \begin{theorem}\label{wloopact}
 The action of the Wilson loop observables $W_{\mu,s}(A)$, $W_{\mu,s}(B)$ on the gauge invariant Hilbert space $\mathcal H_{inv}$  is given by
 \begin{align}
 &\pi(W_{\mu,s}(A))\psi(\alpha,\beta)=e^{is\alpha}\psi(\alpha,\beta-\mu)\qquad \pi(W_{\mu,s})(B)\psi(\alpha,\beta)=e^{is\beta}\psi(\alpha+\mu,\beta)\quad\mu\notin 2\pi\ZZ\nonumber\\
  &\pi(W_{\mu,s}(A))\psi(\alpha,\beta)=\psi(\alpha,\beta-\mu)\qquad\quad\;\; \pi(W_{\mu,s})(B)\psi(\alpha,\beta)=\psi(\alpha+\mu,\beta)\quad\quad\;\; \mu\in 2\pi\ZZ\nonumber.
 \end{align}
It is  a
  unitary  $*$-representation with respect to the star structure \eqref{obsstar} and the canonical  inner product on $\mathcal H_{inv}$ .
 \end{theorem}
 
 It is instructive to consider an infinitesimal version of the quantum observables $W_{\mu,0}(A)$, $W_{\mu,0}(B)$ and variables constructed from the limits $\lim_{\mu\rightarrow 0} W_{\mu,s}(A)$,
 $\lim_{\mu\rightarrow 0} W_{\mu,s}(B)$. We set
 \begin{align}\label{musdef0}
&\hat s_A=-\frac d {d\mu}\bigg|_{\mu=0} \!\!\!\!W_{\mu,0}(A) & &\hat s_B=-\frac d {d\mu}\bigg|_{\mu=0} \!\!\!\!W_{\mu,0}(B)\\
&\hat f(\mu_A)=\sum_{s\in\ZZ/2}\tilde f(s) \lim_{\mu\rightarrow 0} W_{\mu,s}(A) & &\hat f(\mu_B)=\sum_{s\in\ZZ/2}\tilde f(s) \lim_{\mu\rightarrow 0} W_{\mu,s}(B), \nonumber
 \end{align}
 where $f$ is a $2\pi$-periodic function on $\RR$ and $\tilde{f}$ its Fourier transform.
 Using the formulas in Theorem \ref{wloopact}, we find that their action on the gauge invariant Hilbert space $\mathcal H_{inv}$ is given by 
 \begin{align}\label{musact0}
&\pi(\hat s_A)\psi(\alpha,\beta)=\partial_\beta\psi(\alpha,\beta) & &\pi(\hat s_B)\psi(\alpha,\beta)=-\partial_\alpha\psi(\alpha,\beta)\\
 &\pi(\hat f(\mu_A))\psi(\alpha,\beta)=f(\alpha)\cdot \psi(\alpha,\beta) & &\pi(\hat f(\mu_B))\psi(\alpha,\beta)=f(\beta)\cdot \psi(\alpha,\beta).\nonumber
 \end{align}
 The observables $s_A,s_B$ thus act as derivatives, while the observables $\hat f(\mu_A)$, $\hat f(\mu_B)$ act by multiplication. This allows us to directly determine the commutators of the variables $\mu_A,\mu_B, s_A,s_B$, and we obtain the following theorem.
 \begin{theorem} The quantum algebra generated by the observables \eqref{musdef0} is
  isomorphic to two commuting copies of the Heisenberg algebra, which are generated by the pairs of variables $(s_A,\mu_B)$ and $(-s_B,\mu_A)$.
 \end{theorem}
The appearance of the Heisenberg algebra  in the context of combinatorial quantisation is natural, since this quantisation formalism makes use of the Heisenberg double $H(DSU(2))$. The Heisenberg double of a Hopf algebra can be viewed as a generalisation of cotangent bundles and Heisenberg algebras to the context of Hopf algebras. 

The relation between the moduli algebra of gauge invariant quantum observables for the torus universe and the Heisenberg algebra allows one to directly apply results from  quantum mechanics such as uncertainty relations and coherent states to the study of the quantum torus universe. 
Moreover, we will show in  the next section  that the variables $\mu_A,\mu_B$ and $s_A,s_B$ are the quantum counterparts of the classical mass and spin variables introduced in Sect.~\ref{classgeom} and \ref{frsect}. They are thus directly related to the four parameters which characterise the classical geometry of the torus universe.


\section{Classical limit and geometrical interpretation}
\label{btrafo}

The results of the last section provide a precise description of the quantised  torus universe which defines 
 its moduli algebra of gauge invariant  quantum observables and its irreducible representation on the gauge invariant Hilbert space. As it is based on the representation theory of the Drinfel'd double $DSU(2)$, the combinatorial quantisation formalism does not encounter the difficulties arising in a naive implementation of the flatness constraint. Instead,
gauge invariant states and the inner product on the gauge invariant Hilbert space  are obtained naturally from the characters of $DSU(2)$.

However,  the resulting description has the drawback that it obscures the link with the classical theory - Fock and Rosly's Poisson algebra given by \eqref{poissj} to \eqref{poissfj}. It is not readily apparent how the kinematical and gauge invariant  Hilbert space and the associated observables are related to the quantisation of Fock and Rosly's Poisson algebra and its representation \eqref{jfact} on the kinematical Hilbert space $L^2(SU(2)\times SU(2))$.
 In this section, we will establish the relation between the two descriptions. This will allow us to investigate the classical limit of the theory and to clarify the geometrical interpretation of the gauge invariant quantum observables.

\subsection{Decoupling of the Heisenberg double}

The key observation that allows one to relate the two descriptions is the fact that the Heisenberg double $H(DSU(2))$ is isomorphic as an algebra  to the tensor product of two copies of the Heisenberg double $HSU(2)$. As we will see in the following, this isomorphism relates the graph algebra and the algebra  of gauge invariant observables described in Sect.~\ref{combalg} and \ref{combob}
 to the cotangent bundle $T^*(SU(2)\times SU(2))$\ and allows one to make contact with  their classical  description in Sect.~\ref{frsect}.

We start by introducing the Heisenberg double $H(SU(2))$.  
By applying the general definition Def.~\ref{heisdef} in Appendix \ref{hopfapp} to the Hopf algebra $\CC(SU(2))$, we obtain  the following definition.
\begin{definition} \label{hsu2def}The Heisenberg double $H(SU(2))$ is  an associative algebra that as a vector space is isomorphic to the tensor product $\CC(SU(2))\oo \CC(SU(2))^*\cong \CC(SU(2))\oo F(SU(2))$ and 
into which both the group algebra $\CC(SU(2))$ and its dual $F(SU(2))$ are embedded by injective algebra isomorphisms.  In terms of the basis $\{g\oo f\;|\; g\in SU(2), f\in F(SU(2))\}$, its algebra structure is given by
\begin{align}\label{hsu2mult}
(u\oo f)\cdot(v\oo h)=uv\oo h\cdot (f\circ R_v)\qquad u,v\in SU(2), f,g\in F(SU(2)),
\end{align}
where $R_v$ denotes the action of $SU(2)$ on itself by right-multiplication:
$R_v(w)=wv^\inv$ for all  $v,w\in SU(2)$.
It is equipped with a  star structure induced by the star structures \eqref{csu2} and \eqref{fsu2} on $\CC(SU(2))$ and $F(SU(2))$
\begin{align}\label{starhsu}
(u\oo f)^*=u^\inv\oo \bar f \circ R_{u^\inv}.
\end{align}
\end{definition}

We will now show that the  Heisenberg double $H(SU(2))$ is directly related
to the cotangent bundle $T^*SU(2)$. Using the notation
$f(g)\equiv e\oo f(g)$ and $\nabla(u)=u\oo 1$ for $u,g\in SU(2)$, $f\in F(SU(2))$,  we can rewrite the multiplication law  \eqref{hsu2mult} of $H(SU(2))$ as
\begin{align}\label{htang}
(f \cdot h)(v)=f(v)h(v)\quad \nabla(u)\nabla(v)=\nabla(uv)\quad (\nabla(u)\cdot f)(v)=f(vu^\inv)\nabla(u).
\end{align}
If we identify  the variables $f\equiv e\oo f$ with functions on $SU(2)$ and define
\begin{align}\label{nabvec}
&J^i_R=\frac d {dt} |_{t=0} \nabla(e^{-tJ_i})\qquad i=0,1,2
\end{align}
we find that the variables $J^i_R$, $i=0,1,2$ can be identified with the left-invariant vector fields  on $SU(2)$. 
The algebra relations \eqref{htang} then encode the commutativity of functions on $SU(2)$, the action of vector fields on functions $f\in F(SU(2))$ and the Lie bracket of the vector fields.

In Definition \ref{hsu2def}, the  dual  $\CC(SU(2))^*$ of the group algebra $\CC(SU(2))$  is identified with the space $F(SU(2))$ of continuous functions on $SU(2)$. In the following, we relax this restriction and include distributions on $SU(2)$. When generalised to singular elements 
 $u\oo \delta_g$, $u,g \in SU(2)$, expressions \eqref{hsu2mult} and \eqref{starhsu} for the multiplication and star structure then become
 \begin{align}
(u \oo \delta_g)\cdot (v\oo \delta_h) \; = \; uv \oo \delta_h \delta_{gv} \qquad (u \oo \delta_g)^* \; = \; u^{-1} \oo \delta_{gu^\inv} \;.
\end{align}
Using this description in terms of singular functions, one can demonstrate  that the Heisenberg double $H(DSU(2))$ is isomorphic to the tensor product of two copies of $H(SU(2))$. The isomorphism is given by the following lemma.
\begin{lemma} (Decoupling of the Heisenberg double)\label{relpaper}

The Heisenberg double $H(DSU(2))$ is isomorphic as a star algebra to the tensor product of two copies of the Heisenberg double $H(SU(2))$. In terms of the basis $g\oo \delta_u$  of $H(SU(2))$ and the basis $(\delta_u\oo g)\oo (h\oo \delta_v)$ of $H(DSU(2))$, the algebra isomorphism  
$\Phi: H(SU(2))\oo H(SU(2))\rightarrow H(DSU(2))$
is given by
\begin{align}\label{isomorph2}
\Phi((g\oo \delta_u)\oo(h \oo \delta_v)) \; = \; (\delta_{hv^\inv gu^\inv} \oo h)\oo(gu^\inv hu \oo \delta_v) \qquad g,h,u,v\in SU(2).
\end{align}
Its inverse takes the form
\begin{align}\label{inver}
\Phi^\inv((\delta_u\oo g)\oo(h \oo \delta_v)) \; = \; (vg^\inv u g^\inv u^\inv g v^\inv h \oo \delta_{g^\inv u^\inv gv^\inv h}) \oo (g \oo \delta_v) \,.
\end{align}
\end{lemma}

{\bf Proof:} To demonstrate that $\Phi$ defines an algebra isomorphism, it is sufficient to show that 
that $\Phi^\inv \circ \Phi=\text{id}_{HSU(2)\oo HSU(2)}$, $\Phi\circ \Phi^\inv=\text{id}_{H(DSU(2))}$
and that
the elements on the right-hand side of \eqref{isomorph2}  satisfy the multiplication relations of $H(SU(2))\oo H(SU(2))$. The former follows from a direct calculation using expressions \eqref{isomorph2}, \eqref{inver}. The latter
 can be shown by computing the images of the canonical basis of $H(SU(2))\times H(SU(2))$:
\begin{align}\label{isomorph}
\Phi:\;\;&(e\oo \delta_g)\oo (e\oo 1)\mapsto f_1(g)=\int_{SU(2)}\!\!\!\! dv\; (\delta_{v}\oo e)\oo (e\oo \delta_{g^\inv v^\inv})\\
&(e\oo 1)\oo (e\oo \delta_g)\mapsto f_2(g)=(1\oo e)\oo (e\oo \delta_g)\nonumber\\
&(g\oo 1)\oo (e\oo 1)\mapsto \nabla_1(g)=  (1\oo e)\oo (g\oo 1)\nonumber\\
&(e\oo 1)\oo (g\oo 1)\mapsto \nabla_2(g)=\int_{SU(2)\times SU(2)}\!\!\!\!\!\!\!\!\!\!\!\!\!\!\!\!   dvdw\; (\delta_{gv}\oo  g)\oo (wgw^\inv \oo \delta_{wv^\inv})\nonumber,
\end{align} 
where the indices $1$ and $2$ refer to the two commuting $H(SU(2))$ sub-algebras.  From the multiplication law \eqref{hmult} it then follows that the elements $\nabla_i(u)$ and $f_i(v)$ 
satisfy the multiplication relations of $H(SU(2))\oo H(SU(2))$:
\begin{align}\label{isoidhd}
&\nabla_i(g)\cdot f_i(h)=f_i(hg^\inv)\cdot \nabla_i(g) & &\nabla_i(g)\cdot \nabla_j(h)=\nabla_j(h)\cdot \nabla_i(g)\\
&\nabla_i(g)\cdot\nabla_i(h)=\nabla_i(gh)  & &\nabla_i(g)\cdot f_j(h)=f_j(h)\cdot \nabla_i(g)\nonumber\\
&f_i(g)\cdot f_i(h)=f_i(h)\cdot f_i(g)=\delta_h(g)f_i(g)\quad & &f_i(g)\cdot f_j(h)=f_j(h)\cdot f_i(g) \nonumber\quad \text{for}\;i,j=1,2,\;i\neq j.
\end{align}
\hfill$\Box$

As in the case of the Heisenberg double $H(DSU(2))$, the representation theory of the Heisenberg double $H(SU(2))$ is well-understood. It has a single irreducible representation which is given by the following lemma.

\begin{lemma}\label{irrephsu2lemma}
The Heisenberg double $H(SU(2))$ has a single irreducible representation, which is realised on the dual $SU(2)^*$ equipped with the inner product induced by the Haar measure
\begin{align}\label{innsu2*}
\langle \delta_g,\delta_h \rangle=\delta_g(h).
\end{align}
In terms of the singular elements $u\oo \delta_g \in H(SU(2))$ and $\delta_h \in SU(2)^*$ it takes the form
\begin{align}\label{irrephsu2}
\pi_H(u\oo \delta_g)\, \delta_h \; = \; \delta_g(h) \delta_{hu^\inv} \;.
\end{align}
It is  a  $*$-representation with respect to the inner product \eqref{innsu2*} and the star structure \eqref{starhsu}.
\end{lemma}

To obtain explicit expressions for the action of the Heisenberg double $H(SU(2))$ on functions in $F(SU(2))$, we again use an expansion in terms of ``coefficient  functions" \begin{align}
f=\int_{SU(2)}\!\!\!\!\!\!\!\!dz\; f(z)\delta_z.
\end{align}
We  then find that the action of the Heisenberg double $H(SU(2))$ on $F(SU(2))$ is given by
\begin{align}
\pi_H(u\oo \delta_g)f(h)=\delta_{gu^\inv}(h)\,f(hu).
\end{align}
In particular, we have for $u\in SU(2)$ and functions $f,g\in F(SU(2))$
\begin{align}
\pi_H(e\oo g)f(h)=g(h)\cdot  f(h)\qquad \pi_H(u\oo 1)f(h)=f(hu).
\end{align}

\subsection{Kinematical states and classical limit}
\label{kinstatesclass}

To relate the graph algebra to the cotangent bundle $T^*(SU(2)\times SU(2))\cong T^*SU(2)\times T^*SU(2)$, we need to determine the action of  the graph algebra  on $F(SU(2))\oo F(SU(2))\cong F(SU(2)\times SU(2))$.  For this, we apply the isomorphism \eqref{inver} between $H(DSU(2))$ and  $H(SU(2))\otimes H(SU(2))$ to expression \eqref{isoL_HDG} for the generators $A(v,u)$, $B(v,u)$ and obtain
\begin{align}\label{abim}
\Phi^\inv  (A(v,u)) &=   ([u,v]\oo \delta_{v^\inv}) \oo (u \oo 1)\\
\Phi^\inv  (B(v,u)) &=   \int_{SU(2)} \!\!\!\!\! \!\!\!\!\! dw \, (wu^\inv w^\inv \oo \delta_{w^\inv}) \oo (u \oo \delta_{wv^\inv}) \,.\nonumber
\end{align}
Denoting by  $\tilde \pi=(\pi_H\oo \pi_H)\circ \Phi^\inv$ the resulting representation of the graph algebra on $F(SU(2))\oo F(SU(2))$, we find that the action of  the generators $A(v,u)$ and $B(v,u)$ in this representation is given by:
\begin{align}\label{abstate}
&\tilde \pi(A(v,u))(\delta_g\oo \delta_h)=\delta_{v^\inv}(g)\;\delta_{ugu^\inv}\oo \delta_{hu^\inv}\\
&\tilde \pi(B(v,u))(\delta_g\oo \delta_h)=\delta_{v^\inv}(gh)\;\delta_{ug}\oo \delta_{hu^\inv}\nonumber.
\end{align}
The relation between this representation of the graph algebra and the irreducible representation defined in Corollary \ref{graphreps} is given by the following lemma.

\begin{lemma} \label{interlem}The isomorphism $\varphi: DSU(2)^*\rightarrow SU(2)^*\oo SU(2)^*$ defined by
\begin{align}\label{intertw}
\varphi(a\oo \delta_b)=\delta_{b^\inv a^\inv}\oo \delta_b
\end{align}
is an intertwiner between the representations  \eqref{graphrep} and \eqref{abstate} of the graph algebra:
\begin{align}\label{inter}
\tilde \pi(X(v,u))\varphi(a\oo \delta_b)=\varphi(\pi(X(v,u))(a\oo \delta_b))\qquad \forall X\in\mathcal L (T,ISU(2)), a,b\in SU(2).
\end{align}
\end{lemma}

{\bf Proof:} Clearly, $\varphi$ is invertible. To show that it is an intertwiner between the representations \eqref{graphrep} and \eqref{abstate}, it is sufficient to demonstrate that \eqref{inter} holds for the generators $X=A,B$. This follows by a direct computation using  \eqref{graphrep} and \eqref{abstate}.\hfill$\Box$

The representation of the graph algebra on $SU(2)^*\oo SU(2)^*\cong F(SU(2)\times SU(2))$ allows us to make contact with 
 Fock and Rosly's Poisson algebra \eqref{poissj} to \eqref{poissfj} and the associated quantum algebra given by \eqref{jfact}.
 For this, we expand functions $\phi\in F(SU(2)\times SU(2))$ formally in terms of the singular elements $\delta_g\oo \delta_h$
\begin{align}\label{phistatedef}
\phi=\int_{SU(2)\times SU(2)}\!\!\!\!\!\!\!\!\!\!\!\!\!\!\!dgdh\;\; \phi(g,h)\;\delta_g\oo \delta_{g^\inv h}.
\end{align}
Using formulas \eqref{abstate} for the action of  $A(v,u)$, $B(v,u)$ on the singular states $\delta_g\oo \delta_h$, we find that their action on functions $\phi\in F(SU(2)\times SU(2)))$ is given by
\begin{align}\label{obsact}
&\tilde \pi(A(v,u))\phi(g,h)=\delta_{u v^\inv u^\inv }(g)\; \phi(u^\inv g u, [u^\inv,g] h u)\\
&\tilde \pi(B(v,u))\phi(g,h)=\delta_{u v^\inv u^\inv }(h)\; \phi(u^\inv g, u^\inv h u).\nonumber
\end{align}
We  now consider the infinitesimal counterparts $j^a_A$, $j^a_B$  
of the elements $A(v,u)$, $B(v,u)$
\begin{align}\label{jobsdef}
&\hat \j^a_A=-\frac d {dt}\bigg|_{t=0} \int_{SU(2)}\!\!\! \!\!\! \!\!\! dv\; A(v, e^{tJ_a})\qquad &\hat \j^a_B=-\frac d {dt}\bigg|_{t=0} \int_{SU(2)}\!\!\!\!\!\! \!\!\!  dv\; B(v, e^{tJ_a})
\end{align}
as well as observables associated with functions $f\in F(SU(2)\times SU(2))$
\begin{align}\label{fobsdef}
& f =\int_{SU(2)\times SU(2)} \!\!\!\!\! \!\!\!\!\! \!\!\!\!\! dvdw\; f(v^\inv, w^\inv)\; B(w,e)A(v,e).
\end{align}
From \eqref{obsact},  we can determine the action of the observables $j^a_A,j^a_B, \hat f$ on the kinematical states $\phi$ and obtain the following expressions
\begin{align}\label{obsactinf}
&\tilde \pi(\hat \j^a_A)\phi(g,h)=-\frac d {dt}\bigg |_{t=0} \phi(e^{-tJ_a}ge^{tJ_a},[e^{-tJ_a}, g]he^{tJ_a})\\
&\tilde\pi(\hat \j^a_B)\phi(g,h)=-\frac d {dt}\bigg |_{t=0} \phi(e^{-tJ_a}g,e^{-tJ_a}he^{tJ_a})\nonumber\\
&\tilde\pi( f)\phi(g,h)=f(g,h)\cdot \phi(g,h)\nonumber.
\end{align}
This coincides with expressions \eqref{jfact} for the  quantum algebra associated with Fock and Rosly's Poisson structure. Moreover, a direct calculation shows that the commutators of the quantities directly reproduce Fock and Rosly's Poisson brackets \eqref{poissj} to \eqref{poissfj}
\begin{align}
&[\hat \j^a_A,  f](g,h)=-\frac d {dt}\bigg |_{t=0}  \!\!\!\!\! f(e^{-tJ_a}ge^{tJ_a},[e^{-tJ_a}, g]he^{tJ_a}) &
&[\hat \j^a_B, f](g,h)=-\frac d {dt}\bigg |_{t=0}  \!\!\!\!\! f(e^{-tJ_a}g,e^{-tJ_a}he^{tJ_a})\nonumber\\
&[\hat \j^a_A,\hat \j^b_B]=-\epsilon^{ab}_{\;\;\;c}\, \hat \j_A^c & &[\hat \j^a_B,\hat \j^b_B]=-\epsilon^{ab}_{\;\;\;c}\, \hat \j_B^c\nonumber\\
&[\hat \j^a_A,\hat \j^b_B]=-\epsilon^{ab}_{\;\;\;c}\, \hat \j_B^c
& &[f, g]=0.\label{comm}
\end{align}
The observables $\hat \j^a_A$ and $\hat \j_B^a$ defined in \eqref{jobsdef} are thus the quantum counterparts of the vectors $\bj_A,\bj_B\in\RR^3$ which parametrise the translational component of the holonomies in \eqref{ABhol}. The observables  $f$ defined in \eqref{fobsdef} are the quantum counterparts of  functions $f\in F(SU(2)\times SU(2))$ which depend on the rotation component of the two holonomies along the $a$- and $b$-cycle of the torus. This provides a direct physical interpretation of the generators  $A(v,u)$, $B(v,u)$ in terms of the geometry of the torus universe.

In particular, we can determine the action of the Wilson loop observables $W_{\mu,s}(A)$, $W_{\mu,s}(B)$  on the kinematical  states $\phi(g,h)$.  Using the definition \eqref{wdef} of the Wilson loop observables together with the identities in the proof of Lemma  \ref{unlemma} and \eqref{obsact}, we obtain
\begin{align}
\tilde \pi( W_{\mu,s}(A))\phi(g,h)=&\int_{SU(2)}\!\!\!\!\! dz\int_{N_\mu} \!\!\!\!\! dn\; M(g_\mu,n) \; \chi_{N_\mu}^s(n) \delta_{zn^\inv z^\inv}(g)\phi(g, h
z g_\mu z^\inv)\nonumber\\
=\;& e^{is\mu(g^\inv)} \phi(g, h\, e^{\mu \;\hat p(g)^a J_a})\nonumber\\
\tilde \pi(W_{\mu,s}(B))\phi(g,h)=&\int_{SU(2)}\!\!\!\!\! dz\int_{N_\mu} \!\!\!\!\! dn\; M(g_\mu,n) \; \chi_{N_\mu}^s(n) \delta_{zn^\inv z^\inv}(h)
\phi(z g_\mu^\inv z^\inv  g, h)\nonumber\\
=\;& e^{is\mu(h^\inv)} \phi(e^{-\mu\; \hat p(h)^a J_a}  g, h),\label{wloopreps}
\end{align}
where the functions  $\mu, \hat p^a: SU(2)\rightarrow \RR$ given by \eqref{mupdef}. For the associated 
 variables  $\hat s_A$, $\hat s_B$ and  $\hat f(\mu_A)$,  $\hat f(\mu_B)$  defined in \eqref{musdef0} this yields
\begin{align}
&\tilde \pi(\hat s_A)\phi(g,h)= -\frac d {d\mu}\bigg|_{\mu=0}\!\!\!\!\phi(g, h\, e^{\mu \hat p(g)^aJ_a}) &  &\tilde \pi(\hat s_B)\phi(g,h)=-\frac d {d\mu}\bigg|_{\mu=0}\!\!\!\!\phi( e^{-\mu \hat p(h)^aJ_a} g,h)\\
&\tilde \pi(\hat f(\mu_A))\phi(g,h)=f(\mu(g)) \cdot \phi(g,h) & &\tilde \pi(\hat f(\mu_B))\phi(g,h)=f(\mu(h)) \cdot \phi(g,h)\nonumber.
\end{align}
This coincides with expression \eqref{masspinact} for the quantisation of Fock and Rosly's Poisson algebra. The observables $\hat \mu_A,\hat \mu_B$ and $\hat s_A,\hat s_B$ are therefore  the quantum counterparts of the classical mass and spin observables defined in \eqref{ob} which characterise the  geometry of the torus universe. This provides us with a clear geometrical interpretation of the Wilson loop observables $W_{\mu,s}(A),W_{\mu,s}(B)$.

\subsection{The gauge invariant Hilbert space as a regularisation}

We will now demonstrate that  the gauge invariant observables defined in Sect.~\ref{combstates} and \ref{combob} can be viewed as a regularisation of the ill-defined distributional states resulting from a naive implementation of the constraint operators as discussed in Sect.~\ref{constattempt}.

We start by relating the action of the constraint operator  \eqref{constraintC} to the constraint operators \eqref{lconst} and \eqref{tconst} in the quantisation of Fock and Rosly's Poisson algebra. 
By computing the  image  of the constraint
  $C(v,u)$ (\ref{constraintC}) under the  the inverse of the isomorphism  $\Phi: H(SU(2))\otimes H(SU(2))\rightarrow H(DSU(2))$ introduced in  Lemma \ref{relpaper}, we obtain
\begin{align}\label{hdconstr}
\Phi^\inv(C(v,u)) = \int_{SU(2)\times SU(2)} \!\!\!\!\!\!\!\!\!\!\!\!\!\!\! \!\!\!\!\!da\, db\;\; \delta_{uvu^\inv}([a,b]) \, \left( [b^\inv,u^\inv] \otimes \delta_{u^\inv b u}\right)\! \otimes\! \left( [a^\inv,u^\inv] \otimes \delta_{u^\inv a u}\right).
\end{align}
Using formula \eqref{irrephsu2} for the irreducible representations of the Heisenberg double $H(SU(2))$ and the expansion \eqref{phistatedef},
we  find that the action of the constraint operator on the coefficient functions takes the form
\begin{align}\label{cstates}
\tilde\pi(C(v,u))\phi(g,h)=\delta_{uvu^\inv}([g^\inv,h]) \phi(u^\inv g u, u^\inv h u).
\end{align}
If we define in analogy to \eqref{jobsdef} and \eqref{fobsdef}
\begin{align}
\hat\j_C^a=-\frac d {dt}\bigg|_{t=0} \int_{SU(2)}\!\!\!\! \!\!\!\! dv\; C(v, e^{tJ_a})\qquad \hat f= \int_{SU(2)}\!\!\!\! \!\!\!\! dv\; f(v) C(v,e),
\end{align}
we recover the action  \eqref{lconst}, \eqref{tconst} of the constraint operators on the quantum algebra associated with Fock and Rosly's Poisson algebra and the Poisson bracket \eqref{constcl} of the classical  constraint variables
\begin{align}
&\tilde \pi(\hat \j_C^a)\phi(g,h)\!=\!-\frac d {dt}\bigg |_{t=0} \!\!\!\!\!\!\phi(e^{-tJ_a}g  e^{tJ_a}\!,\! e^{-tJ_a} h  e^{tJ_a})\!=\!\{j^a_C,\phi\}(g,h)\qquad
\tilde \pi(\hat f)\phi(g,h)\!=\!f([g^\inv\!,h]) \phi(g,h).\nonumber
\end{align}
This allows us to compare the implementation of the constraint in the combinatorial quantisation formalism in Sect.~\ref{physhilb} to the naive implementation of constraints  in Sect.~\ref{constattempt}.
For this we determine 
the image of the gauge invariant states \eqref{FTstates} under the intertwiner \eqref{intertw}, which relates the representations of  the graph algebra on $DSU(2)^*$ to the kinematical states $\phi(g,h)$:
\begin{align}\label{stateim}
\varphi(\psi)
=&\int_{SU(2)}\!\!\!\!dz\int_0^{2\pi}\!\!\!\!\!d\alpha\int_0^{2\pi}\!\!\!\!\!d\beta\; \psi(\alpha,\alpha+\beta)\; \delta_{zg_\alpha^\inv z^\inv}\oo \delta_{z g_\beta^\inv z^\inv}\;.
\end{align}
Comparing this expression with the expansion \eqref{phistatedef},
we find that the states $\varphi(\psi)$ are characterised by ``coefficient functions" 
\begin{align}
\phi(g,h)=\delta_e([g,h]) \phi_0(g,h)\qquad\text{with}\quad \phi_0(g,h)=\psi(\mu_g, \mu_h)
\end{align}
where $\psi\in L^2([0,2\pi]\times [0,2\pi])$ and  $\mu_g=\mu(g^\inv)$, $\mu_h=\mu(h^\inv)$  are given by \eqref{mupdef}.  The inner product of two distributional states
\begin{align}
\phi(g,h)=\delta_e([g,h])\,\psi(\mu_g,\mu_h)\qquad \xi(g,h)=\delta_e([g,h])\, \kappa(\mu_g,\mu_h)\qquad \phi,\kappa\in \mathcal H_{inv},\nonumber
\end{align}
 is given by
 the standard inner product on $L^2([0,2\pi]\times  [0,2\pi])$
\begin{align}
\langle \phi, \xi\rangle=\int_0^{2\pi}\!\!\!d\mu_g\int_0^{2\pi}\!\!\!d\mu_h\; \overline{ \psi(\mu_g,\mu_h)}\kappa(\mu_g,\mu_h).
\end{align}
This provides a precise definition of the gauge invariant states which gives a meaning to the formal expression \eqref{naivestate} that results  from a naive implementation of the constraints: States of the form \eqref{naivestate} are to be interpreted as elements of the Hilbert space $L^2([0,2\pi]\times [0,2\pi])$ and their inner product \eqref{naiveprod} is to be identified with the  inner product on $L^2([0,2\pi]\times [0,2\pi])$. 

By  using  the representation theory of the Drinfel'd double $DSU(2)$ as a central ingredient, the combinatorial quantisation scheme thus overcomes the difficulties associated with a naive implementation of the constraints via Dirac's constraint quantisation formalism. It leads to a precise definition of the gauge invariant Hilbert space in terms of $L^2$-functions  on the torus and to explicit expressions for the action of the gauge invariant observables on this space.

\section{The action of the mapping class group}
\label{mapsec}

We conclude our discussion of the quantised torus universe with an investigation of the role of mapping class groups as symmetries in the quantum theory. Mapping class group symmetries in classical and quantised 3d gravity are of conceptual importance in quantum gravity because they are related to the action of large, i.~e.~not infinitesimally generated, diffeomorphisms. As large diffeomorphisms are not generated  by constraints in the classical theory, it has been debated if they should be considered as gauge or as physical symmetries of the theory \cite{giu1,giu2}. Moreover, it has been found  in  several quantisation approaches to 3d gravity that imposing invariance  under the action of the relevant mapping class group
leads to undesirable properties in the resulting quantum theories \cite{giu3, peld1, peld2, carliptor1,carliptor2}.

For spacetimes of topology $\RR\times \Sigma$, where $\Sigma$ is an oriented surface of general genus and with a general number of punctures, elements of the mapping class group $\text{Map}(\Sigma)$ are in one-to-one correspondence with equivalence classes of large modulo small (infinitesimally generated) diffeomorphisms. The mapping class group $\text{Map}(\Sigma)$ is isomorphic to the quotient \begin{align}
\text{Map}(\Sigma)=\text{Out}(\pi_1(\Sigma))/\text{Inn}(\pi_1(\Sigma)). 
\end{align}
of outer modulo inner automorphisms of the fundamental group $\pi_1(\Sigma)$ \cite{birmgreen, birmorange}. 

 There are two mapping class groups which are relevant for the combinatorial quantisation of the torus universe. The first is the mapping class group $\text{Map}(T\setminus D)$ of the punctured torus. It acts on the fundamental group $\pi_1(T\setminus D)$, which is the free group with two generators
$\pi_1(T\setminus D)=\langle a,b\rangle=F_2$.
This action on the $\pi_1(T\setminus D)$ gives rise to an action of the mapping class group $\text{Map}(T\setminus D)$ on the kinematical Hilbert space, i.~e.~the representation space
of the graph algebra defined in Corollary \ref{graphreps}, and on the graph algebra. 
The second is the mapping class group $\text{Map}(T)$ of the torus, the modular group $SL(2,\ZZ)$,
which acts on the fundamental group of the torus $\pi_1(T)=\langle a,b\,;\, ba^\inv b^\inv a=1\rangle\cong \ZZ\times \ZZ$. As we will see in the following, this action gives rise to an action of modular group  on the 
 gauge invariant Hilbert space in Def.~\ref{hinvth} and on the moduli algebra of gauge invariant quantum observables.

\subsection{The modular group and the mapping class group of the punctured torus}

Both, the mapping class group $\text{Map}(T\setminus D)$ of the punctured torus and the modular group $\text{Map}(T)\cong SL(2,\ZZ)$
are generated by 
 Dehn twists along the $a$- and $b$-cycle of the (punctured) torus  \cite{birmgreen, birmorange, cox, rank}.  
 The action of these Dehn twists  on the fundamental groups $\pi_1(T\setminus D)$ and $\pi_1(T)$ is given by
\begin{align}\label{dtdef}
&D_a:\;a\mapsto a &  &D_b:\; a\mapsto b^\inv \cdot a\\
&\qquad\;\, b\mapsto b\cdot a & &\qquad\;\, b\mapsto b.\nonumber
\end{align}
Although both mapping class groups can be presented in terms of the generators $D_a$, $D_b$, they differ with respect to  their defining relations. In the case of the torus, the generators $a$ and $b$ of the fundamental group $\pi_1(T)$ commute, which they do not in the fundamental group $\pi_1(T\setminus D)$ of the punctured torus $T\setminus D$. This leads to additional relations in the presentation of the mapping class group $\text{Map}(T)$, which are not present in $\text{Map}(T\setminus D)$. In particular, all inner automorphisms of $\pi_1(T)\cong \ZZ\times \ZZ$ are trivial, and its mapping class group is  given by $\text{Out}(\pi_1(T))$.

The mapping class group  $\text{Map}(T)$ of the torus is isomorphic to the modular group $SL(2,\ZZ)$ \cite{birmorange, cox, rank}. The relation of this description to the description in terms of Dehn twists  in \eqref{dtdef} is obtained from  the following $GL(2,\ZZ)$ action on $\pi_1(T)\cong \ZZ\times \ZZ$
\begin{align}\label{modact}
\left(\begin{array}{cc} c & d\\ e & f\end{array}\right):\; (a,b)\mapsto( b^e \cdot a^f, b^c\cdot a^d)\qquad c,d,e,f\in \ZZ. 
\end{align}
The restriction of this action to the subgroup $SL(2,\ZZ)\subset GL(2,\ZZ)$ defines the action of the modular group on the fundamental group $\pi_1(T)\cong \ZZ\times \ZZ$. In this formulation, the generating Dehn twists \eqref{dtdef} are represented by the matrices
\begin{align}\label{dtmat}
M_A=\left(\begin{array}{cc} 1 & 1\\ 0 & 1\end{array}\right)\qquad M_B=\left(\begin{array}{cc} 1 & 0\\-1 & 1\end{array}\right).
\end{align}
To make contact with the standard presentation of $SL(2,\ZZ)$ in terms of generators and relations, we consider the elements $T=D_a$ and $S=D_a^\inv \circ D_b^\inv\circ D_a^\inv$. From \eqref{dtdef}  we then find that the element $S$ acts on the fundamental group $\pi_1(T)$ according to
\begin{align}\label{sact}
S:\; a\mapsto b,\; b\mapsto a^\inv. 
\end{align}
The representing $SL(2,\ZZ)$ matrices associated to $S$ and $T$ are therefore given by
\begin{align}\label{stdef}
T=\left(\begin{array}{cc} 1 & 1\\ 0 & 1\end{array}\right)\qquad S=\left(\begin{array}{cc} 0 & -1\\ 1 & 0\end{array}\right),
\end{align}
which are the standard generators  of the modular group \cite{cox,rank}. They generate the modular group $SL(2,\ZZ)$ subject to the defining relations
\begin{align}\label{strel}
(ST)^3=S^2\qquad S^4=1.
\end{align}
In the construction of the torus as a quotient of $\RR^2$  in Sect.~\ref{classgeom}, the action of the modular group  corresponds to a change of the fundamental parallelograms  in Fig.~\ref{torus}. The generators $a,b\in\pi_1(T)$ which define the tessellation of the plane by parallelograms change non-trivially.   From this change of the parallelograms, one can derive the transformation of the moduli $\tau \in\CC$  introduced in \eqref{taudef} under the action of the modular group. One finds that the  generators $T,S$ in \eqref{stdef} act on the moduli   according to\begin{align}
S: \tau\mapsto -\frac 1 \tau\qquad\qquad T: \tau\mapsto \tau+1.
\end{align}
Although the fundamental parallelograms and hence the moduli of the torus change under the action of the modular group,
the associated action of $\pi_1(T)\cong \ZZ\times\ZZ$ on $\RR^2$ and hence the metric on the resulting torus do not change. The action of the modular group thus relates two different but equivalent descriptions of the same physical state.

\subsection{Mapping class actions in the quantised torus universe}

We will now investigate the role of mapping class groups in the combinatorial quantisation of the Euclidean torus universe. We start by considering the kinematical Hilbert space $L^2(SU(2)\times SU(2))$, i.~e.~the representation space of the graph algebra defined in Corollary \ref{graphreps}. As the quantum flatness constraint \eqref{constraintC}, which corresponds to the defining relation of the fundamental group $\pi_1(T)$, is not imposed on this Hilbert space, the relevant mapping class group is the mapping class group $\text{Map}(T\setminus D)$ of the punctured torus. 

A detailed investigation of the action of mapping class groups on Fock and Rosly's Poisson structure \cite{FR} and the associated quantum algebra is given in \cite{we3} for an orientable surface $\Sigma$   of genus $g\geq 0$ and with $n\geq 0$ punctures. 
 By specialising the results of \cite{we3} to the torus, one obtains the following theorem.
\begin{theorem} \cite{we3} 
\begin{enumerate}
\item The mapping class group $\text{Map}(T\setminus D)$ acts on Fock and Rosly's Poisson algebra  \eqref{poissj} by Poisson isomorphisms.  
\item The kinematical Hilbert space $L^2(SU(2)\times SU(2))$  carries a unitary representation of the mapping class group $\text{Map}(T\setminus D)$, which is given by  \begin{align}
 \Gamma\phi(g,h)=\phi\circ \rho_\Gamma(g,h),
\end{align}
 where $\Gamma\in\text{Map}(T\setminus D)$ and $\rho_\Gamma: SU(2)\times SU(2)\rightarrow SU(2)\times SU(2)$ is the action of $\text{Map}(T\setminus D)$ induced by its action on $\pi_1(T\setminus D)$. 
 \item The action of the generating Dehn twists \eqref{dtdef} is given by
 \begin{align}
 D_a\phi(g,h)=\phi(g,h g)\qquad\qquad D_b\phi(g,h)=\phi(h^\inv g,h).\label{dtkin}
 \end{align}
\end{enumerate}
\end{theorem}
It is instructive to compare expressions \eqref{dtkin} for the action of the generators $D_a,D_b$ on $L^2(SU(2)\times SU(2))$ with formula \eqref{obsact} for the representation of the elements $A(v,u), B(v,u)$. One finds that the action of the Dehn twists $D_a,D_b$ on $L^2(SU(2)\times SU(2))$ coincides with the representation of the following elements of the graph algebra
\begin{align}
D_a\phi=\tilde\pi\left(\int_{SU(2)}\!\!\!\! \!\!\!\! du \;A(u^\inv,u)\right)\phi\qquad D_b\phi=\tilde\pi\left(\int_{SU(2)}\!\!\!\! \!\!\!\! du\; B(u^\inv,u)\right)\phi.
\end{align}
It remains to relate the action of the mapping class group $\text{Map}(T\setminus D)$ to the  gauge invariant observables of the theory, the fundamental Wilson loop observables $W_{\mu,s}(A)$, $W_{\mu,s}(B)$. For this purpose, we define for each element $X$ of the graph algebra an associated  observable $O_X$ 
\begin{align}\label{xmapdef}
O_X=&\frac 1 {2\pi}\sum_{s\in\ZZ/2}\int_0^{2\pi} \!\!\!\!d\mu\; \; e^{i\mu s}\;W_{\mu,s}(X).
\end{align}
The representation of the observables $O_A$, $O_B$ on the kinematical Hilbert space  is then obtained from formula \eqref{wloopreps} for the action of the Wilson loop observables $W_{\mu,s}(A)$, $W_{\mu,s}(B)$:
\begin{align}\label{obs1}
\tilde \pi(O_A)\phi(g,h)=&\frac 1 {2\pi}\sum_{s\in\ZZ/2}\int_0^{2\pi}\!\!\!\!d\mu\;e^{is\mu} e^{is\mu(g^\inv)}\phi(g, h\cdot e^{\mu \hat p(g)^a J_a})\\
\tilde \pi(O_B)\phi(g,h)=&\frac 1 {2\pi}\sum_{s\in\ZZ/2}\int_0^{2\pi}\!\!\!\!d\mu\;e^{is\mu} e^{is\mu(h^\inv)}\phi(e^{-\mu \hat p(h^\inv)^a J_a} g, h)\nonumber
\end{align}
To evaluate this expression further, we perform the summation over the variable $s$, which yields a  $2\pi$-periodic delta distribution. The expressions \eqref{obs1} reduce to
\begin{align}
&\tilde \pi(O_A)\phi(g,h)=\int_0^{2\pi}\!\!\!\!d\mu\; \delta_{2\pi}(\mu(g^\inv)+\mu)\,\phi(g, h\cdot e^{\mu \hat p(g)^a J_a})=\phi(g,hg)\\
&\tilde \pi(O_B)\phi(g,h)=\int_0^{2\pi}\!\!\!\!d\mu\; \delta_{2\pi}(\mu(h^\inv)+\mu)\,\phi(e^{-\mu \hat p(h^\inv)^a J_a} g, h)=\phi(h^\inv g,h).\nonumber
\end{align}
By comparing this expression with formula \eqref{dtkin} for the action of the generating Dehn twists, we obtain the following theorem.
\begin{theorem}
The action of the generating Dehn twists \eqref{dtkin} on the kinematical Hilbert space $L^2(SU(2)\times SU(2))$ is given by  the representation of the observables $O_A,O_B$ 
\begin{align}
D_a\phi=\tilde \pi(O_A)\phi\qquad D_b\phi=\tilde\pi(O_B)\phi\qquad\forall \phi\in L^2(SU(2)\times SU(2)).
\end{align}
\end{theorem}

To conclude our discussion, we investigate  the action of the modular  group $\text{Map}(T)\cong SL(2,\ZZ)$  on the  gauge invariant Hilbert space
$\mathcal H_{inv}$. For this we note that the identification of the two arguments with the generators of the fundamental group $\pi_1(T)=\ZZ\times \ZZ$ induces an action of the modular group on $\mathcal H_{inv}$. By comparing this action to the action of the observables $O_A,O_B$, we obtain the following theorem.

\begin{theorem}
The modular group $SL(2,\ZZ)$ acts unitarily on the gauge invariant Hilbert space $\mathcal H_{inv}$ defined in Theorem \ref{hinvth} according to
\begin{align}\label{mapstate}
\left(\begin{array}{cc} c & d\\ e & f\end{array}\right)\psi(\alpha,\beta)=\psi(f\alpha+e\beta, d\alpha+c \beta)\qquad \forall \left(\begin{array}{cc} c & d\\ e & f\end{array}\right)\in SL(2,\ZZ).
\end{align}
The action of the generating Dehn twists \eqref{dtdef} via the representing matrices \eqref{dtmat} coincides with the representation of the Wilson loop observables $O_A, O_B$ defined in \eqref{xmapdef}:
\begin{align}
\tilde \pi(O_A)\psi(\alpha,\beta)=M_A\psi(\alpha,\beta)\qquad \tilde \pi(O_B)\psi(\alpha,\beta)=M_B\psi(\alpha,\beta)\qquad\forall \psi\in \mathcal H_{inv}.\nonumber
\end{align}
\end{theorem}

{\bf Proof:} The fact that \eqref{mapstate} defines a representation of  $GL(2,\ZZ)$ on $\mathcal H_{inv}$ follows by direct calculation from the multiplication law of $GL(2,\ZZ)$ and formula \eqref{mapstate}. Its restriction to $SL(2,\ZZ)$ thus defines a representation of the modular group. To show that this representation preserves the inner product \eqref{FTinnprod},
we note that the Jacobi determinant of transformation \eqref{mapstate} coincides with the determinant of the associated $SL(2,\ZZ)$ matrix. 

To show that the action of the generating Dehn twists \eqref{dtdef} coincides with the action of the Wilson loop observables $O_A, O_B$, we use the formulas for the action of the Wilson loop observables $W_{\mu,s}(A)$, $W_{\mu,s}(B)$ on $\mathcal H_{inv}$ from Theorem \ref{wloopact}
together with the definition \eqref{xmapdef} of the observables  $O_X$. This yields
\begin{align}
&\tilde \pi(O_A)\psi(\alpha,\beta)\!=\!\frac 1 {2\pi}\!\sum_{s\in\ZZ/2} \int_0^{2\pi}\!\!\!\!d\mu\; e^{is(\alpha+\mu)}\psi(\alpha,\beta\!-\!\mu)\!=\!\int_0^{2\pi}\!\!\!\!d\mu\; \delta_{2\pi}(\mu\!+\!\alpha) \psi(\alpha,\beta\!-\!\mu)\!=\!\psi(\alpha,\beta+\alpha)\nonumber\\
&\tilde \pi(O_B)\psi(\alpha,\beta)\!=\!\frac 1 {2\pi}\!\sum_{s\in\ZZ/2} \int_0^{2\pi}\!\!\!\!d\mu\; e^{is(\beta\!+\!\mu)}\psi(\alpha+\mu,\beta)\!=\!\int_0^{2\pi}\!\!\!\!d\mu\; \delta_{2\pi}(\mu+\beta) \psi(\alpha\!+\!\mu,\beta)\!=\!\psi(\alpha-\beta,\beta).\nonumber
\end{align}
\hfill $\Box$

The representation  of the modular group on the gauge invariant Hilbert space $\mathcal H_{inv}$ induces an action of the modular group on the algebra  $\mathcal A_{inv}$  which is generated by the Wilson loop observables $W_{\mu,s}(A)$ and $W_{\mu,s}(B)$.
As a general element $\Gamma$ of the modular group  can be expressed as a product of the generators $D_a,D_b$ and their inverses
\begin{align}
\Gamma=D_a^{\alpha_1} D_b^{\beta_1}D_a^{\alpha_2}D_b^{\beta_2}\cdots D_a^{\alpha_n}D_b^{\beta_n}\qquad\text{with}\; \alpha_i,\beta_i\in \ZZ, 
\end{align}
its action on a Wilson loop observable is given by
\begin{align}
\Gamma W_{\mu,s}(X)=\left(O_A^{\alpha_1} O_B^{\beta_1}\cdots O_A^{\alpha_n}O_B^{\beta_n} \right) \,W_{\mu,s} (X)\left(O_A^{\alpha_1} O_B^{\beta_1}\cdots O_A^{\alpha_n}O_B^{\beta_n}\right)^*,
\end{align}
where  $X=A,B$ and $*$ denotes   the star structure \eqref{obsstar}.

By definition, the matrix elements of observables
are invariant under a simultaneous action of the modular group on states and observables. 
However, imposing invariance of the gauge invariant states under the action of the modular group destroys the structure of the gauge invariant Hilbert space. By \eqref{mapstate}, invariance of a state $\psi\in \mathcal H_{inv}$ under the action of the modular group  amounts to the condition
\begin{align}
\psi(a\alpha+b\beta, c\alpha+d\beta)=\psi(\alpha,\beta)\qquad\forall a,b,c,d\in\ZZ, ad-bc=1.
\end{align}
As there exist elements of the modular group for which the action  on the torus is ergodic \cite{brown}, any continuous function satisfying this condition must be constant. This is incompatible with the definition of the gauge invariant Hilbert space as the set of $L^2$-functions on the torus.  

This result is in agreement with  the problems encountered in implementing modular  invariance for the  quantum torus universe in  \cite{giu3, peld1,peld2,carliptor1,carliptor2}. Although these papers are based on very different quantisation approaches, in all of them the requirement that the physical states of the theory are invariant under the action of the modular group leads to problems and ambiguities in the definition of the Hilbert space. 

It would be interesting to understand how our result is related to the descriptions in \cite{giu3, peld1,peld2,carliptor1,carliptor2} and to identify a common source of this problem. This would require relating the timeless formulation in terms of holonomies to time dependent quantisation formalisms.  However, even without such an explicit relation, the fact that  imposing modular invariance of the physical states 
appears to be problematic in all quantisation approaches
 suggests that it may be too strong a requirement.


\section{Outlook and Conclusions}
\label{outlook}

In this article, we extended the combinatorial quantisation formalism for Chern-Simons theories with compact, semisimple gauge groups \cite{AGSI,AGSII,AS,BR} to the Euclidean torus universe with the non-compact and non-semisimple gauge group $ISU(2)$. The representation theory of the associated quantum group, the Drinfel'd double $DSU(2)$, gives rise to considerable complications in this  approach. As its representation spaces are infinite-dimensional and labelled by continuous parameters,
the definition of traces and of the character ring, which play a central role in this formalism, 
becomes problematic and requires a careful discussion of measures and distributions.

We showed how these problems arising from the representation theory of $DSU(2)$  can be addressed and obtained a rigourous Hamiltonian quantisation formalism for the Euclidean torus universe based on the quantum group symmetries of the theory. In this formalism,  the gauge invariant Hilbert space is identified with the space  square integrable functions on the torus, and the quantum algebra of gauge invariant observables is directly related to two commuting copies of the Heisenberg algebra. 
We demonstrated that
the gauge invariant Hilbert space carries a unitary representation of the modular group, which is given by the action of certain Wilson loop observables associated with the $a$- and $b$-cycle of the torus.  As combinatorial quantisation is based on an explicit correspondence between classical and quantum observables, the resulting description allows one to directly perform the classical limit of the theory and to obtain a clear interpretation of the quantum observables in terms of the parameters that characterise the classical geometry of the torus universe.

The description  is based on similar variables and closely related to other approaches to quantum gravity such as the loop formalism in three dimensions and the Ponzano-Regge model \cite{MN1}. However, the fundamental difference is that it makes use of quantum group symmetries as a central ingredient in the quantisation of the theory and in the implementation of the constraints. These symmetries  lead to a precise definition of the gauge invariant Hilbert space  that resolves the problems and ambiguities arising from a naive implementation of the constraints in the loop formalism such as in \cite{NP}. 

As the resulting Hilbert space is given by the square integrable functions on the torus and the algebra of gauge invariant observables is given by  two commuting copies of the Heisenberg algebra, our quantisation formalism allows one to directly investigate spectra of observables, uncertainty relations and coherent states  by adapting results from quantum mechanics. 
In particular, it would be interesting to investigate the spectra of  observables with a direct geometrical interpretation such as the area of the torus universe  and the lengths of its $a$- and $b$-cycle for a given value of the ``time" parameter $t$. A comparison with the results  and conclusions of \cite{spectra} which are based on an extended (non-gauge invariant) Hilbert space could shed  light on the debate surrounding the relation between the spectra of gauge invariant and kinematical observables \cite{thombianc, rovspec}.

It would also be interesting to establish a relation with previous work on the (Lorentzian) torus universe \cite{ giu3, peld1,peld2, carnel, carnel2, carliptor1,carliptor2, carliptor3}. This would require
relating  our holonomy-based formalism to  time-dependent formalisms along the lines of \cite{carnel,carnel2,carliptor2}. The relation between the holonomies and the moduli of the torus should then allow one to compare the resulting descriptions. In particular, it would be interesting to determine the impact of the signature (Euclidean vs Lorentzian) and the cosmological constant.

We expect that our quantisation formalism can be generalised to Chern-Simons theories 
with  gauge groups $G\ltimes\mathfrak g$ on manifolds of topology $\RR\times \Sigma$, where $\Sigma$ is an oriented surface of genus $g\geq 0$ and with $n\geq 0$ punctures and $G$ a unimodular Lie group with Lie algebra $\mathfrak g$. The generalisation to gauge groups $G\ltimes\mathfrak g$ and topology $\RR\times T$ should be direct. 
In that case, the relevant quantum group would be the Drinfel'd double $DG$, whose representation theory is investigated in detail in  \cite{kM, KBM}. 
In particular, this would include the  Lorentzian torus universe with $G=SO(2,1)$.
The generalisation to surfaces
of higher genus and/or with punctures can be expected to be more challenging. Although it would
pose no problems  with respect to the representation theory,  the combinatorics of the resulting description would become considerably more involved. 

\section*{Acknowledgements}
C.M.'s work is funded by the German Research Foundation (DFG) via the Emmy Noether fellowship  ME 3425/1-1. She is also a member of the Collaborative Research Center 676 ``Particles, Strings and the
Early Universe". K.N.'s work is partially supported by the ANR.

\appendix

\section{Hopf algebra structures}
\label{hopfapp}

\subsection{The group algebra $\mathbb C(SU(2))$ and its dual}

The group algebra $\CC(SU(2))$ has a canonical Hopf $*$-algebra structure, which is most easily formulated in a basis consisting of all group elements   $g\in SU(2)$. In this basis, the Hopf algebra operations are given by 
\begin{align}\label{csu2}
&\text{multiplication:}\; \cdot(a\oo b)=a\cdot b &
&\text{unit:}\; \eta(z)=z\cdot e & &\text{antipode:}\; S(g)=g^\inv\\
&\text{comultiplication:}\; \Delta(a)=a\oo a &
&\text{counit:}\; \epsilon(g)=1 & &\text{$*$-structure:}\; g^*=g^\inv.\nonumber
\end{align}
Its dual Hopf algebra  can be identified with the algebra $F(SU(2))$ of functions on $SU(2)$. The pairing between $\CC(SU(2))$ and $F(SU(2))$ is  given by $\langle g, f \rangle=f(g)$ for $g\in G, f\in F(SU(2))$.
The Hopf algebra structure on $F(SU(2))$  can be derived from the one on $SU(2)$ via  the principle of Hopf algebra duality. It takes the form
\begin{align}\label{fsu2}
&\text{multiplication:}\; (f\cdot_* h)(u)=f(u)h(u) &
&\text{unit:}\; \eta_*(u)=1 & &\text{$*$-structure}: f^*=\bar f\\
&\text{comultiplication:}\; \Delta_*(f)(u, v)=f(u\cdot v) &
&\text{counit:}\; \epsilon_*(f)=f(e) &
&\text{antipode:}\; S_*(f)(u)=f(u^\inv),\nonumber
\end{align}
where $f,h\in F(SU(2))$, $u, v\in SU(2)$.

\subsection{Heisenberg doubles}
The Heisenberg double $H(\mathcal A)$  of  a Hopf algebra $\mathcal A$  was first defined and investigated \cite{AF,STS}. It is 
 an associative algebra ({\em not} a Hopf algebra) which can be viewed as a generalisation of the cotangent bundle $T^*G$ of a Lie group $G$ to the context of Hopf algebras. It can be characterised as the minimal (non-trivial) associative algebra into which both the Hopf algebra $\mathcal A$ and its dual $\mathcal A^*$ can be embedded via injective algebra homomorphisms. We have the following general definition.

\begin{definition} (Heisenberg double \cite{AF,STS})\label{heisdef}

The {\em Heisenberg double} $H(\mathcal A)$ of a Hopf algebra $\mathcal A$ is the vector space $\mathcal A\oo\mathcal A^*$ equipped with the unique structure of an associative algebra such that the inclusions $i_{\mathcal A}: \mathcal A\rightarrow H(\mathcal A), x\mapsto x\oo 1$ and $i_{\mathcal A^*}: \mathcal A^*\rightarrow H(\mathcal A), \alpha\mapsto 1\oo \alpha$ are injective algebra homomorphisms. In terms of the basis $x\oo \alpha$, $x\in\mathcal A$, $\alpha\in \mathcal A^*$ the algebra multiplication is given by
\begin{align}
&(x\oo \eta_*)\cdot (y\oo \eta_{*})=(x\cdot y)\oo \eta_*\qquad \forall x,y\in\mathcal A\\
&(\eta\oo \alpha)\cdot(\eta\oo \beta)=\eta\oo (\alpha\cdot_*\beta)\qquad\forall \alpha,\beta\in\mathcal A^*\\
&(x\oo \eta_*)\cdot (\eta\oo \alpha)=x\oo \alpha=\sum_{(x), (\alpha)} \langle x_{(1)}\, \alpha_{(2)}\rangle\; (\eta\oo \alpha)\cdot(x\oo \eta_{*}),
\end{align}
where $x,y\in\mathcal A$, $\alpha,\beta\in\mathcal A^*$ and we used Sweedler's notation $\Delta(x)=\sum_{(x)} x_{(1)}\oo x_{(2)}\in \mathcal A\oo \mathcal A$, $\Delta_*(\alpha)=\sum_{(\alpha)} \alpha_{(1)}\oo \alpha_{(2)}\in\mathcal A^*\oo \mathcal A^*$.
\end{definition}

The representation theory of  Heisenberg doubles was first investigated in \cite{AF}. It is shown there that the Heisenberg double of a Hopf algebra $\mathcal A$ has a single irreducible representation realised on its dual $\mathcal A^*$. 

\begin{theorem} \cite{AF}\label{rephdth}

The Heisenberg double Hopf algebra  $H(\mathcal A)$  of a Hopf algebra $\mathcal A$ admits a unique irreducible representation. This representation is realised on the dual Hopf algebra $\mathcal A^*$  and is given by
\begin{align}\label{hdrepgen}
&\pi: \;H(\mathcal A)\rightarrow \text{End}(\mathcal A^*)\\
&\pi(x)\alpha=(1\oo \langle\, x,\,\cdot\;\rangle)\circ\Delta_*(\alpha)\qquad \pi(\beta)\alpha=\beta\cdot_* \alpha\qquad\forall x\in\mathcal A, \alpha,\beta\in\mathcal A^*,\nonumber
\end{align}
where $\cdot_*: \mathcal A^*\oo \mathcal A^*\rightarrow \mathcal A^*$ and $\Delta_*: \mathcal A^*\rightarrow \mathcal A^*\oo \mathcal A^*$ denote, respectively, the multiplication and comultiplication of $\mathcal A^*$ and $\langle\,,\rangle: \mathcal A\oo \mathcal A^*\rightarrow \CC$ the pairing between $\mathcal A$ and $\mathcal A^*$.
\end{theorem}

\section{Representation theory of $SU(2)$}
\label{su2char}

\subsection{Wigner functions}

We denote by $\rho_J: SU(2)\rightarrow \text{End}(V_J)$ the irreducible unitary representations of $SU(2)$, labelled by half-integers $J\in\NN_0/2$ and of dimension $d_J=\text{dim}_\CC(V_J)=2J+1$.
As a consequence of the Peter Weyl theorem, a dense orthonormal basis of the space of functions on $SU(2)$ is provided by its matrix elements in the irreducible unitary representations $\rho_I$ or, equivalently, by the Wigner functions
\begin{align}\label{wigdef}
D^J_{jm}(g)={\sqrt{d_J}} \pi_J(g)_{jm}={\sqrt{d_J}}\langle J,j\,|\, \rho_I(g) J,m \rangle\qquad j,m\in\{-J, -J+1 ,..., J\}.
\end{align}
They satisfy the condition
\begin{align}\label{jident}
D^J_{jm}(g \cdot e^{\mu J_0})=e^{im\mu}D^J_{jm}(g)\quad \forall g\in SU(2),\; J_0=\tfrac 1 2\left(\begin{array}{cc} i & 0 \\ 0 & -i\end{array}\right).
\end{align}
The fact that the Wigner functions form a dense orthonormal basis of the space of functions on $SU(2)$ 
is encoded in the orthogonality relations
\begin{align}\label{wigorth}
\int_{SU(2)}\!\!\!\!dz\; \overline{D^J_{jm}(z)}D^{I}_{kl}(z)= \delta_{I,J}\delta_{jk}\delta_{ml}
\end{align}
where $dz$ denotes the Haar measure of $SU(2)$, and in their completeness relations
\begin{align} \label{wigcomp}
\sum_{J\in\NN_0/2}\sum_{k,l\in\{-J,...,J\}} \overline{D^J_{kl}}(g)D^J_{kl}(h)=\sum_{J\in\NN_0/2} \text{Tr}_J(D^J(gh^\inv))=\delta_h(g),
\end{align}
where $\text{Tr}_J$ is the trace in the representation labelled by $J\in\NN_0/2$ and
 $\delta_h$ denotes the Dirac delta distribution with respect to the Haar measure on $SU(2)$.

\subsection{Characters}

The  characters  $\chi_J: SU(2) \rightarrow \RR$ of the
unitary irreducible representations $\rho_J:SU(2)\rightarrow \text{End}(V_J)$ labelled by $J\in\NN_0/2$
 are given by
\begin{align}
\chi_J(g)=\text{Tr}(\rho_J(g))\qquad \forall g\in SU(2).
\end{align}
They satisfy
\begin{align}\label{charexpl}
\chi_J(e^{\alpha J_0})=\sum_{k=0}^{J} e^{i(J-k)\alpha}=\frac{\sin{((2J+1)\alpha/2)}}{\sin (\alpha/2)}\qquad \text{where}\qquad J_0=\tfrac 1 2 \left(\begin{array}{cc} i & 0 \\ 0 & -i\end{array}\right).
\end{align}
The characters $\chi_J$, $J\in\NN_0/2$,  form a dense  orthonormal basis of the space of class functions on $SU(2)$. This follows from  their orthogonality relations 
\begin{align}\label{orthoirr}
\int_{SU(2)}  \chi_I(g)\chi_J(g) dg=\delta_{I,J}
\end{align}
and their completeness relations
\begin{align}\label{comp}
\sum_J \chi_J(g)\chi_J(h)=\delta_g(h),
\end{align}
where $\delta_g$ denotes the Dirac delta distribution  on the space of class functions on $SU(2)$.

\end{document}